\documentclass[nofootinbib,aps,amssymb,floatfix,superscriptaddress,preprintnumbers]{revtex4}

\usepackage{graphicx}
\usepackage{bm}
\usepackage{amsmath}
\usepackage{amssymb}
\usepackage{amsfonts}
\usepackage{float}
\usepackage[colorlinks,pdfusetitle]{hyperref}
\hypersetup{colorlinks=true,allcolors=[rgb]{0,0.0,1.0}}
\usepackage{dsfont}  
\usepackage{slashed}  
\usepackage{booktabs}
\usepackage{multirow}
\usepackage{subfigure}
\usepackage[sort&compress]{natbib}
\usepackage{xcolor}
\usepackage{ulem}
\usepackage[percent]{overpic}

\newcommand{\be}{\begin{equation}}  
\newcommand{\ee}{\end{equation}}  
\newcommand{\beq}{\begin{eqnarray}} 
\newcommand{\eeq}{\end{eqnarray}}

\newcommand{\nn}{\nonumber \\}

\newcounter{RSQ}

\begin{document}

\title{ Probing quantum entanglement with Generalized Parton Distributions\\ at the Electron-Ion Collider }

\author{Yoshitaka Hatta  }

\author{Jakob Schoenleber}
\affiliation{Physics Department, Brookhaven National Laboratory, Upton, NY 11973, USA}
\affiliation{RIKEN BNL Research Center, Brookhaven National Laboratory, Upton, NY 11973, USA}

\begin{abstract}
Within the collinear factorization framework based on Generalized Parton Distributions (GPDs), we calculate the spin density matrix of  exclusively produced quark and antiquark pairs $u\bar{u}$, $d\bar{d}$, $s\bar{s}$,  $c\bar{c}$,   $b\bar{b}$ in  electron-proton scattering. The presence of both real and imaginary parts in the scattering amplitudes leads to a rich pattern of entanglement between the quark and the antiquark. 
We  map out kinematical regions where the pairs exhibit   entanglement, Bell nonlocality and non-stabilizerness (`magic'). We also predict that massive  quarks and antiquarks are transversely polarized, similar to the well-known transverse hyperon polarization in unpolarized collisions. In   strangeness, charm and bottom productions, the polarization can reach 50-80\% in certain kinematic regions in the low-energy runs of the Electron-Ion Collider. 
\end{abstract}

\maketitle

\section{Introduction}

Entanglement is a quintessential feature of quantum mechanics that requires  the coherent description of a composite quantum system as a whole.  
While it manifests across  vastly different scales in both system size  and particle number,  historically  
stringent experimental tests of entanglement have primarily focused on the spin or  polarization  of  elementary particle pairs. A classic example is the entanglement of photon polarizations in the  (para-)positronium decay ${\rm Ps}\to 2\gamma$ \cite{Wu:1950zz}. More   definitive  tests including the violation of Bell's inequality \cite{Bell:1964kc}, or the Clauser-Horne-Shimony-Holt (CHSH) inequality also involve two-photon final states   \cite{Clauser:1969ny,Aspect:1981zz,Weihs:1998gy}. Photons are advantageous because their polarizations can be measured relatively easily in table-top  or small-scale laboratory experiments. 

Entanglement is also ubiquitous  among the other elementary particles of the Standard Model that are  produced in  large-scale collider experiments. However, its direct  measurement is much more challenging for a number of reasons. First, many redundant particles are produced in high energy collisions and they may interfere with the  particles of interest. Besides, if the strong interaction is involved,   entanglement generated at the partonic level can be easily washed out by the hadronization process. Moreover, the polarization of  final state particles is  often difficult to measure.   Under these  circumstances, in practice, one can only infer entanglement between a pair of heavy, short-lived  particles by measuring the angular distribution of their weak decay products. The first experimental confirmation of entanglement in  the top-antitop ($t\bar{t}$) system, as predicted by theory   \cite{Bernreuther:1993hq,Baumgart:2012ay,Bernreuther:2015yna,Afik:2020onf,Dong:2023xiw}, has been achieved by  the ATLAS and CMS collaborations at the LHC   \cite{ATLAS:2023fsd,CMS:2024pts,CMS:2024zkc} (see also an earlier attempt \cite{CMS:2019nrx}). Expectations are high 
\cite{Barr:2024djo,Afik:2025ejh} that  similar measurements can be done for other particle species  such as $W^+W^-$ \cite{Aguilar-Saavedra:2022mpg,Fabbrichesi:2023cev}, $\tau^+\tau^-$ \cite{Altakach:2022ywa,Han:2025ewp} and $b\bar{b}$ \cite{Kats:2023zxb,Afik:2025grr} pairs, etc.,  and even light quark-antiquark pairs   \cite{Cheng:2025cuv}. Moreover,  a violation of the Bell-CHSH inequality, or `Bell nonlocality' could also be detected in these systems \cite{Fabbrichesi:2021npl,Severi:2021cnj,Afik:2025ejh}, although it adds another layer of practical  challenges.\footnote{  The recent controversy (e.g., \cite{Abel:1992kz,Abel:2025skj,Low:2025aqq}) prompts us to comment that, by Bell nonlocality, we mean  certain properties of spin density matrices  computed within QCD (see Section IV.B). It is not meant to be another test of quantum mechanics or the (non-)existence of local hidden variable theories \cite{Low:2025aqq}.  }

In this paper, we continue the discussion of spin-spin entanglement between quark-antiquark ($q\bar{q}$) pairs produced in  exclusive Deep Inelastic Scattering (DIS) and Ultraperipheral Collisions (UPCs) along the lines of  \cite{Qi:2025onf,Fucilla:2025kit}. Our work is particularly motivated by the future  high-luminosity DIS experiment at the Electron-Ion Collider (EIC) \cite{AbdulKhalek:2021gbh}. Since an electron is involved in the initial state, DIS offers a cleaner (i.e., less produced particles) environment for studying entanglement than in hadron-hadron collisions. One can realize an even cleaner environment, akin to the final states of $e^+e^-$ annihilation, by focusing on exclusive (diffractive) processes   
where   the target proton does not break up.   In \cite{Fucilla:2025kit}, the spin density matrix of this process has been computed in the  Regge limit where the collision energy is asymptotically large and only the color-singlet gluonic  exchange (`Pomeron') needs to be  considered. It has been found that the $q\bar{q}$ pairs are always entangled and always exhibit  Bell nonlocality. On the other hand, the EIC runs with variable electron-proton $e+p$ center-of-mass (CM) energies in the range $28<\sqrt{s}<140$ GeV which only partially overlaps with the Regge regime.  A solid theoretical framework to describe exclusive processes in this window  is the Generalized Parton Distribution (GPD) \cite{Diehl:2003ny,Belitsky:2005qn}. So far, the study of GPDs has been mostly centered around the goal of obtaining the three dimensional tomographic picture of the nucleon. Here, for the first time, we propose to use GPDs as a tool to probe quantum entanglement at the EIC.  
We will be particularly interested in whether entanglement and Bell nonlocality are  generic features of color singlet exchanges in QCD, or they are emergent phenomena in the high energy limit. We also evaluate the so-called `magic' \cite{Leone:2021rzd,White:2024nuc}, a necessary nonclassical resource for achieving quantum advantage in quantum computing.

As a byproduct, we find nonvanishing polarization of massive $q\bar{q}$ pairs, even though the colliding particles are unpolarized. This is a kind of single spin asymmetry, and is analogous to  the transverse polarization of hyperons in unpolarized proton-proton collisions and  semi-inclusive DIS  \cite{Bunce:1976yb,Lundberg:1989hw,HERMES:2007fpi,COMPASS:2021bws}. We demonstrate that the polarization can be quite sizable, exceeding 60\% in certain kinematical regions.

\section{Exclusive dijet cross section} 

In this section, we  review the exclusive dijet cross section  calculated  in the collinear factorization  framework  \cite{Braun:2005rg}. This sets the stage for the calculation of the associated spin density matrix in the next section. 
To lowest order, `dijet' simply means an energetic quark-antiquark ($q\bar{q}$) pair. `Exclusive' means that that the target proton scatters elastically. Thus the process of interest is  $\gamma^{(*)}+p \to q+\bar{q}+p'$.  This  can be thought of as a subprocess in DIS where the virtual photon $\gamma^*$ with virtuality $Q^2=-q^2$ is emitted from the incoming electron. In proton-nucleus ($p+A$) UPCs, the photon $\gamma$ is real ($Q^2\approx 0$), emitted from the incoming heavy nucleus. We work in a frame where the photon is  collinear\footnote{The light-cone coordinates are defined as $p^\mu=(p^+,p^-,{\bm p})$ with $p^\pm=\frac{1}{\sqrt{2}}(p^0\pm p^3)$. Boldface letters denote two-dimensional vectors ${\bm p}=(p^1,p^2)$ and we use $i,j=1,2$ for their indices. } $q^\mu = (q^+,-\frac{Q^2}{2q^+},{\bm 0})$ and fast-moving $q^+ \gg Q$ in the $+x^3$ direction. The incoming and outgoing protons have momenta 
\beq
p^\mu \approx \left( 0,(1+\xi)P^-, -\frac{{\bm \Delta}}{2} \right), \qquad p'^\mu \approx\left( 0,(1-\xi)P^-, \frac{{\bm \Delta}}{2} \right).
\eeq
We will neglect the proton mass and assume ${\bf \Delta}\approx 0$ throughout this paper. The latter approximation avoids possible entanglement between the $q\bar{q}$ pair and the proton in momentum space. Moreover, the proton spin does not affect the spin state of the pair because, at high energy, the proton helicity is conserved in near-forward scattering.  

In this setup, the produced $q\bar{q}$ pair has momentum 
\beq
\tilde{k}^\mu = \left(zq^+, \frac{k_\perp^2+m^2}{2zq^+},{\bm k}\right), \qquad 
\tilde{k}'^\mu = \left(\bar{z}q^+,\frac{k_\perp^2+m^2}{2\bar{z} q^+},-{\bm k}\right) ,\label{fast} 
\eeq
and invariant mass 
\beq
M^2=(\tilde{k}+\tilde{k}')^2 = \frac{k_\perp^2+m^2}{z\bar{z}}.
\eeq
$0<z=\frac{p\cdot \tilde{k}}{p\cdot q}<1$ is the fraction of the photon energy carried by the quark ($\bar{z}=1-z$ for the antiquark). One may regard this process as the  $2\to 2$ scattering  $\gamma^{(*)}+{\mathbb P}\to q+\bar{q}$ where ${\mathbb P}$ denotes `Pomeron'\footnote{The use of the word `Pomeron' is due to a lack of better terminology. In the present context, it  simply means the  color singlet two-gluon or quark-antiquark exchange in the $t$-channel. In the high energy (Regge) limit, it is smoothly connected to the Pomeron exchange  discussed in \cite{Fucilla:2025kit}.}  with momentum 
\beq
{\mathbb P}^\mu = p^\mu - p'^\mu \approx \left(0,2\xi P^-, {\bm 0}\right).
\eeq
The skewness variable $\xi$ is fixed by momentum conservation in this $2\to 2$ scattering as  
\beq
\frac{2\xi}{1+\xi} = \frac{M^2+Q^2}{W^2+Q^2} = \frac{M^2+Q^2}{ys}, \label{skew}
\eeq
with $W^2=(p+q)^2$ being the $\gamma^*+p$ center-of-mass (CM) energy. On the right hand side, we introduced the  $e+p$ CM energy $s$ and the standard DIS variable $y=\frac{q\cdot p}{l\cdot p}$.
 
We write the unpolarized cross section as 
\beq
\frac{d\sigma^{ep}}{dW^2dQ^2dz d^2{\bm k}d^2{\bm \Delta}}=\frac{\alpha_{em}}{ \pi sQ^2}  \frac{1+(1-y)^2}{2y} \left( \frac{d \sigma^{ T}}{d z  d^2 \boldsymbol{k}d^2{\bm \Delta} } +\varepsilon \frac{d \sigma^{ L}}{d z  d^2 \boldsymbol{k}d^2{\bm \Delta} } \right)+\cdots. \label{ep}
\eeq
  The superscripts $T$ and $L$ denote the contributions from the transversely and longitudinally polarized virtual photons and $\varepsilon = \frac{1-y}{1-y+\frac{y^2}{2}}$ is the ratio of their fluxes. The terms omitted in (\ref{ep}) are proportional to $\cos\phi$ or $\cos 2\phi$ where $\phi$ is the azimuthal angle of ${\bm k}$ relative to the lepton scattering plane. These terms arise from interference between different photon polarization states. We neglect them assuming an implicit integration over  $\phi_k$. 
In Ultraperipheral Collisions (UPCs), only the transverse part is relevant 
\beq
\frac{d\sigma^{\rm UPC}}{dzd^2{\bm k}d^2{\bm \Delta}}= \int dq^0 \frac{dN}{dq^0} \frac{d\sigma^T}{dzd^2{\bm k}d^2{\bm \Delta}},
\eeq
where $dN/dq^0$ is the well-known equivalent photon flux  \cite{Jackson:1998nia}. 

In \cite{Braun:2005rg}, $d\sigma^{L/T}$ has been calculated  to leading order in collinear factorization using the GPDs.  The result reads 
\beq
    \frac{d \sigma^L}{d z  d^2 \boldsymbol{k} d^2{\bm \Delta} }&=&  \frac{\alpha_{em}e_q^2z^2\bar{z}^2Q^2\alpha_s^2}{2\pi^2N_c (1-\xi^2)(k_\perp^2+\mu^2)^4}|I_L^g+2C_FI_L^q|^2 , 
\label{crossL}\\
 \frac{d \sigma^T}{d z  d^2 \boldsymbol{k} d^2{\bm \Delta} } &=& \frac{\alpha_{em}e_q^2\alpha_s^2}{16\pi^2 N_c(1-\xi^2)(k_\perp^2+\mu^2)^4}  \nn && \quad \times \biggl( k_\perp^2\Bigl(\bigl|2C_F(I_T^{q_1}+I_T^{q_2})-(1-2z)I_T^g\bigr|^2+\bigl|2C_F(I_T^{q_1}-I_T^{q_2})+I_T^g\bigr|^2\Bigr)+2m^2|I_L^g|^2\biggr),   \label{crossT}
\eeq
where  $\alpha_{em}=\frac{1}{137}$, $\alpha_s=\frac{g^2}{4\pi}$ is the QCD coupling,  and $\mu^2\equiv m^2+z\bar{z}Q^2$.  $I$'s are convolution integrals with GPDs 
\beq
I_L^g&=& \int_{-1}^1 dx F_g(x,\xi)\left( \frac{4\xi (1-\beta)(x^2+\xi^2)}{(x^2-\xi^2+i\epsilon)^2} - \frac{2\xi (1-2\beta)}{x^2-\xi^2+i\epsilon}\right) \nn
I_L^q &=& \int_{-1}^1 dx F_q(x,\xi) \left( \frac{2\xi \bar{z}}{x+\xi-i\epsilon} + \frac{2\xi z}{x-\xi+i\epsilon}\right),\nn
I_T^g &=& \int_{-1}^1 dx F_g(x,\xi) \left( \frac{2\xi(1-2\beta) (x^2+\xi^2)}{(x^2-\xi^2+i\epsilon)^2} +\frac{4\xi \beta}{x^2-\xi^2+i\epsilon} \right), 
\label{I}
\\
I_T^{q_1}&=&\int_{-1}^1 dx F_q(x,\xi) \left( \frac{2\xi z\bar{z}}{x-\xi+i\epsilon} -\frac{2\xi \beta \bar{z}^2}{(1-\beta)(x+\xi-i\epsilon)} + \frac{2\xi \bar{z}^2}{(1-\beta)(x-\xi(1-2\beta)-i\epsilon)}\right),
\nn
I_T^{q_2}&=&\int_{-1}^1 dx F_q(x,\xi) \left( \frac{2\xi\beta z^2}{(1-\beta)(x-\xi+i\epsilon)} -\frac{2\xi z\bar{z}}{x+\xi-i\epsilon} - \frac{2\xi z^2}{(1-\beta)(x+\xi(1-2\beta)+i\epsilon)}\right). \notag
\eeq
$I_L^q,I_T^{q_{1,2}}$ and  $I_L^g,I_T^g$ come from the $t$-channel quark-antiquark and two-gluon exchange diagrams, respectively. In (\ref{I}), we introduced  the variable  
\beq
\beta= \frac{\mu^2}{k_\perp^2+\mu^2} , 
\eeq
often used in diffractive DIS and defined 
\beq
F_{q,g} \equiv H_{q,g}-\frac{\xi^2}{1-\xi^2}E_{q,g}. \label{nonflip}
\eeq 
The quark and gluon GPDs $H_q$ and $H_g$ are normalized such that they reduce to the respective PDFs $H_q(x,0)=q(x)$ and  $H_g(x,0)=xg(x)$ in the forward limit. On the other hand, the GPDs $E_q,E_g$ have no collinear analogs. 
 The linear combination (\ref{nonflip}) represents the helicity nonflip amplitude \cite{Diehl:2003ny}.  
In the following, we are only concerned with the region $\xi\ll 1$, and accordingly neglect ${\cal O}(\xi^2)$ terms. The GPD $E$'s  drop out in this approximation. 

\begin{figure}
    \centering
\includegraphics[width=1\linewidth]{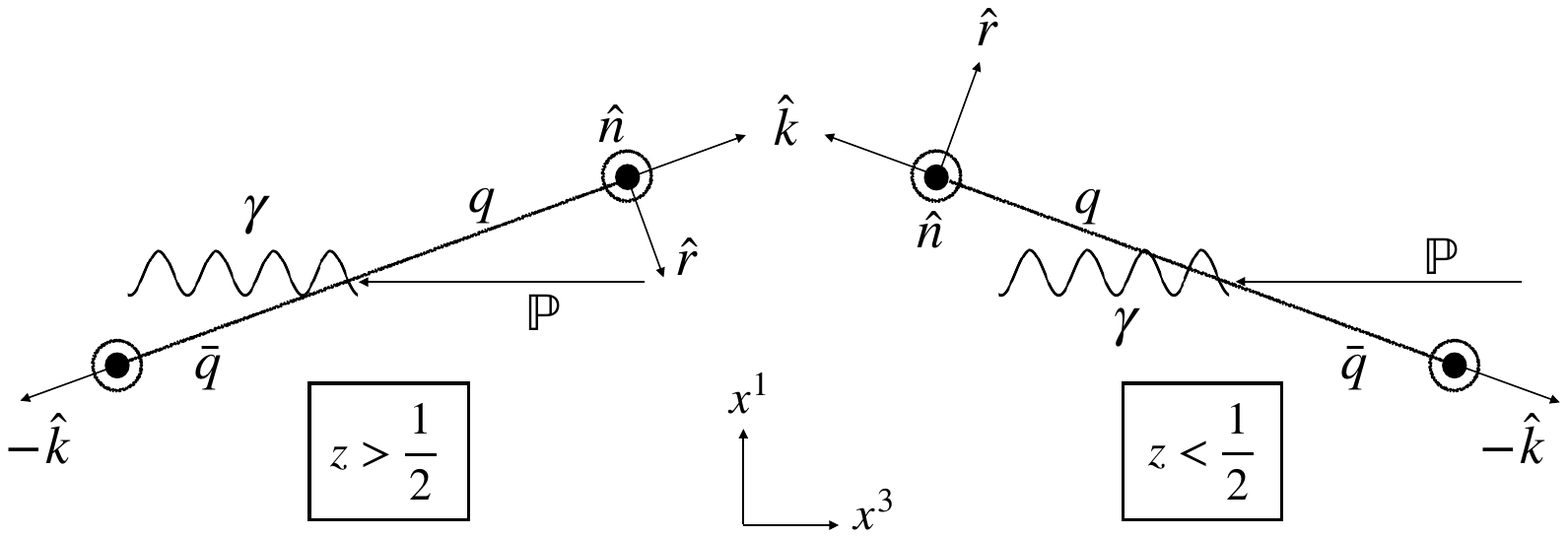}
\vspace{-70mm}
    \caption{$q\bar{q}$ production in the photon-Pomeron CM frame at $\phi=0$. Left: $z>\frac{1}{2}$, Right: $z<\frac{1}{2}$. }
    \label{kinematics}
\end{figure}

\section{Spin density matrix}

The cross sections (\ref{crossL}), (\ref{crossT})  have been obtained by summing over the two spin states of the produced quark and antiquark. In order to compute the spin density matrix, one has to undo this step. Let $\alpha,\alpha'$ be the spin indices of the quark and the  antiquark in the amplitude, and $\beta,\beta'$ be those in the complex-conjugate amplitude.  Before setting $\alpha=\beta$,  $\alpha'=\beta'$ and  summing over $\alpha,\alpha'$,  the squared scattering amplitude
takes the generic form 
\beq
\bar{u}_{\alpha} ( \tilde{k} ) \Gamma v_{\alpha'} (\tilde{k}')\bar{v}_{\beta'} ( \tilde{k}' ) \gamma^0\Gamma^\dagger \gamma^0 u_{\beta} ( \tilde{k})  . \label{uuuu}
\eeq
Up to a prefactor, this is essentially the spin density matrix  with matrix indices  $\alpha\alpha'\beta\beta'$ in spin space. It describes the quantum state of two `qubits,' representing the spin states of the quark and antiquark with fixed momenta.  
Clearly, the density matrix is frame-dependent because spin and momentum do not transform independently  under Lorentz boosts. Results in different frames are  related by Wigner rotations. It also depends on the choice of the spinor basis. On the other hand, since Wigner rotations are unitary,  physical results such as entanglement and Bell-nonlocality are frame-independent \cite{Peres:2002wx,Barr:2024djo}. It is then most convenient to calculate spin density matrix using the so-called Jacob-Wick helicity states \cite{Jacob:1959at} and in the back-to-back frame of the $q\bar{q}$ pair.

\subsection{Longitudinal photon}

In the longitudinal case, (\ref{uuuu}) takes the form  \cite{Braun:2005rg}
\beq
\bar{u}_{\alpha} ( \tilde{k} ) \gamma^+ v_{\alpha'} (\tilde{k}')\bar{v}_{\beta'} ( \tilde{k}' ) \gamma^+ u_{\beta} ( \tilde{k})  . \label{uuuuL}
\eeq
This is exactly the same as in the Pomeron exchange contribution  calculated in \cite{Fucilla:2025kit}.  
We then immediately see that the spin density matrix $C_{ab}^L$  is the same. 
It came as a surprise in \cite{Fucilla:2025kit} that $C_{ab}^L$   resulting from the Pomeron exchange is identical to that for the lowest order process $\gamma^*_L+g\to q+\bar{q}$ calculated in \cite{Qi:2025onf}. That is, $C_{ab}^L$ does not depend on whether  color singlet or  octet states are exchanged in the $t$-channel, or whether one gluon or multiple gluons are exchanged. We now see that $C^L_{ab}$ remains the same even after including the quark-antiquark exchange channel.

For completeness, and to set up our notation for the transversely polarized case to be discussed later, here we reproduce the result (see the appendix of \cite{Fucilla:2025kit}).  First we  go to the photon-Pomeron CM frame where the quark and the antiquark have  back-to-back momenta 
(switching to the Cartesian coordinates $k^\mu =(k^0,\vec{k})= (k^0,{\bm k},k^3)$)
\beq 
k^\mu=
 \left(\frac{M}{2}, {\bm k
 }, \frac{M}{2}(z-\bar{z})\right) \equiv (k^0,\vec{k}), \qquad 
k'^\mu=
\left(\frac{M}{2}, -{\bm k}, -\frac{M}{2}(z-\bar{z})\right) =(k^0,-\vec{k}),
\eeq
and velocity
\beq
|\vec{k}|=\frac{M}{2}v, \qquad v = \sqrt{1-\frac{4m^2}{M^2}}.
\eeq
This can be achieved via a Lorentz boost along the $x^3$ axis $\tilde{k}^\pm= e^{\pm\eta}k^\pm$, $\tilde{k}'^\pm= e^{\pm\eta}k'^\pm$ with 
\beq
e^\eta
= q^+\sqrt{\frac{2z\bar{z}}{k_\perp^2+m^2}}.
\eeq 
In this frame, the quark moves in the direction $\hat{k}=\frac{\vec{k}}{|\vec{k}|}=(\sin\theta \cos\phi, \sin\theta \sin \phi, \cos\theta)$ with polar angles 
\beq
\cos\theta = \frac{(z-\bar{z})M}{\sqrt{M^2-4m^2}}, \qquad \phi={\rm arg}(k^1+ik^2). \label{cos}
\eeq
Note that $\cos\theta\to 1$ when $z\to 1$ and $\cos\theta\to -1$ when $z\to 0$.  Under this boost, 
the spinor product transforms as  
\beq
\bar{u}_{\alpha} ( \tilde{k} ) \gamma^+ v_{\alpha'} (\tilde{k}') =e^\eta\bar{u}_{\alpha} (k ) \gamma^+ v_{\alpha'} (k') .
\eeq
We then perform a spatial rotation from the $(\hat{x}^1,\hat{x}^2,\hat{x}^3)$ frame to the $(\hat{n},\hat{r},\hat{k})$ frame where (see Fig.~\ref{kinematics})
\beq
\hat{n} = \frac{\hat{x}^3\times \hat{k}}{\sin \theta} = (-\sin \phi, \cos\phi,0), \qquad 
\hat{r}= \hat{k}\times \hat{n} =
(-\cos\theta\cos\phi,-\cos\theta \sin \phi, \sin \theta).
\eeq
Under the same rotation, $\gamma^+$ transforms as 
\beq
\gamma^+ = \frac{1}{\sqrt{2}}\left(\gamma^0 +\sin \theta \gamma^r+\cos\theta \gamma^k\right).
\eeq

The spin density matrix is conveniently calculated in the $(\hat{n},\hat{r},\hat{k})$ frame with $\hat{k}$ being the spin quantization axis.  In this frame, the Jacob-Wick spinors take the form    
\beq
u_{\alpha}(k)=\begin{pmatrix} \sqrt{k\cdot \sigma}\xi_\alpha \\ \sqrt{k\cdot \bar{\sigma}}\xi_\alpha \end{pmatrix}, 
\qquad v_{\alpha'}(k')= \begin{pmatrix} \sqrt{k'\cdot \sigma}\tilde{\eta}_{-\alpha'} \\ -\sqrt{k'\cdot \bar{\sigma}}\tilde{\eta}_{-\alpha'} \end{pmatrix} = \begin{pmatrix} \sqrt{k\cdot \bar{\sigma}}\tilde{\eta}_{-\alpha'} \\ -\sqrt{k\cdot \sigma}\tilde{\eta}_{-\alpha'} \end{pmatrix}, \label{heli}
\eeq
with ($\sigma^k=\sigma^3$) and 
\beq
\sqrt{k\cdot \sigma}
= \sqrt{\frac{M}{8}}\left( \sqrt{1+v}(1-\sigma^k)+\sqrt{1-v}(1+\sigma^k)\right), \qquad  
\sqrt{k\cdot \bar{\sigma}} = \sqrt{\frac{M}{8}}\left( \sqrt{1-v}(1-\sigma^k)+\sqrt{1+v}(1+\sigma^k)\right).
\eeq
Here,  $\alpha,\alpha'=\pm$ refers to the helicity (angular momentum projection $\pm \frac{1}{2}$ along the direction of motion). For the antiquark moving in the $-\hat{k}$ direction, we employ the `flipped spinors' 
$\tilde{\eta}_{-\alpha'}=-i\sigma^2 (\eta_{\alpha'})^*$  \cite{Peskin:1995ev}.    
Using these basis, we can write the cross section in the form
\beq
\frac{d\sigma^L}{dzd^2{\bm k}d^2{\bm \Delta}} = \sum_{\alpha=\beta, \alpha'=\beta'}A^L \xi^\dagger_\alpha \eta^\dagger_{\alpha'} 
\rho^L_{\alpha\beta,\alpha'\beta'
}\xi_\beta\eta_{\beta'}, \label{lsigma}
\eeq
where 
\begin{equation}
    A^L= \frac{\alpha_{em}e_q^2\alpha_s^2z^2\bar{z}^2Q^2}{2\pi^2N_c (1-\xi^2)(k_\perp^2+\mu^2)^4}|I_L^g+2C_FI_L^q|^2 ,
\end{equation}
is the unpolarized cross section (\ref{crossL}). 
The density matrix $\rho^L$ can be  parametrized by  
\beq
\rho^L=\frac{1}{4}\left( {\mathds 1}\otimes {\mathds 1} +B_a^{L} \sigma^a\otimes {\mathds 1}+\bar{B}^{L}_b {\mathds 1}\otimes  \sigma^b+ C_{ab}^L\sigma^a \otimes \sigma^b \right), \qquad (a,b=n,r,k) \label{paramet}
\eeq
where ${\mathds 1}$ is the unit $2\times 2$ matrix and $\sigma^{a}$ are the Pauli matrices. 
The vectors $B^L_a$, $\bar{B}^L_b$ represent the polarization of the quark and the antiquark, 
\beq
B^L_a = {\rm Tr}[(\sigma^a\otimes {\mathds 1})\rho^L ], \qquad \bar{B}_b^L= {\rm Tr}[({\mathds 1}\otimes \sigma^b)\rho^L],
\eeq
while the matrix $C^L_{ab}$ describes the correlation between the quark and antiquark spins.  

 In the actual computation, we evaluate (\ref{uuuuL}) using  (\ref{heli}) and  the explicit form of the gamma matrices $\gamma^0=\begin{pmatrix} 0 & 1 \\ 1 & 0\end{pmatrix}$, $\vec{\gamma}=\begin{pmatrix} 0 & \vec{\sigma} \\ -\vec{\sigma} & 0 \end{pmatrix}$. 
We then exploit the fact that any 2 $\times$ 2 matrix can be expanded in the basis $\{ 1, \sigma^n, \sigma^r, \sigma^k \}$ 
\beq
\xi_\beta \xi^\dagger_{\alpha} &=& \frac{1}{2} \left( \xi^\dagger_\alpha \xi_\beta + (\xi^\dagger_\alpha \sigma^n \xi_\beta) \sigma^n + (\xi^\dagger_\alpha \sigma^r \xi_\beta) \sigma^r +(\xi^\dagger_\alpha \sigma^k \xi_\beta) \sigma^k \right) \; , \nn 
\tilde{\eta}_{-\alpha'} \tilde{\eta}^\dagger_{-\beta'} &=& \frac{1}{2} \left( \tilde{\eta}^\dagger_{-\beta'} \tilde{\eta}_{-\alpha'} + (\tilde{\eta}^\dagger_{-\beta'} \sigma^n \tilde{\eta}_{-\alpha'}) \sigma^n + (\tilde{\eta}^\dagger_{-\beta'} \sigma^r \tilde{\eta}_{-\alpha'}) \sigma^r +(\tilde{\eta}^\dagger_{-\beta'} \sigma^k \tilde{\eta}_{-\alpha'}) \sigma^k \right) \; . \label{xixi}
\eeq
Finally we use the formula $
\tilde{\eta}_{-\beta'}^\dagger \vec{\sigma}\tilde{\eta}_{-\alpha'} =
-\eta^\dagger_{\alpha'}\vec{\sigma}\eta_{\beta'}$
to recast the density matrix in the desired form.  
The result is $B^{L}_a=\bar{B}_b^{L}=0$ and 
\beq
C_{nn}^L = 1, &\qquad& 
C_{rr}^L=-C_{kk}^L = -\frac{1-(2-v^2)\cos^2\theta}{1-v^2\cos^2\theta} = -\frac{k_\perp^2-(1-2z)^2m^2}{k_\perp^2+(1-2z)^2m^2}, \nn  && C_{rk}^L=C_{kr}^L = -\frac{\sqrt{1-v^2}\sin 2\theta}{1-v^2\cos^2\theta} = \frac{2(1-2z)k_\perp m}{k_\perp^2+(1-2z)^2m^2}.
\label{cij}
\eeq
The other components are zero. Note that $(C_{rr}^L)^2+(C_{rk}^L)^2=1$.

The matrix $C_{ab}^L$ describes a two-qubit state that is maximally entangled and moreover maximally violates the Bell-CHSH inequality \cite{Qi:2025onf}. In particular, for the symmetric pair $z=\frac{1}{2}$ ($\cos\theta=0$), or in the relativistic limit $k_\perp\to \infty$, or in the massless case $m=0$, $C^L={\rm diag}(1,-1,1)$  represents one of the Bell states 
\beq
\rho^L=|\Phi^+\rangle \langle \Phi^+|, \qquad |\Phi^+\rangle = \frac{1}{\sqrt{2}} \bigl(|++\rangle +|--\rangle\bigr).\label{bell1}
\eeq
where $\pm$ refers to spin projection $\pm \frac{1}{2}$ along the $\hat{k}$-axis. (This is opposite in sign to helicity for the antiquark moving in the $-\hat{k}$ direction.) 
On the other hand, in the forward limit $k_\perp\to 0$ ($\cos\theta=\pm 1$), $C^L={\rm diag}(1,1,-1)$ represents another Bell state
\beq
\rho^L=|\Psi^+\rangle \langle \Psi^+|, \qquad |\Psi^+\rangle = \frac{1}{\sqrt{2}} \bigl(|+-\rangle +|-+\rangle\bigr).  \label{bell2}
\eeq
Moreover, along the line  $z=\frac{1}{2}\pm \frac{k_\perp}{2m}$, 
\beq
C^L =\begin{pmatrix} 1 & 0 & 0 \\ 0 &0 & \mp 1 \\ 0 & \mp 1 & 0\end{pmatrix} 
\eeq
This represents a state equivalent to the  Bell state up to a local phase rotation 
\beq
\rho^L=|\Phi_n^\pm\rangle \langle \Phi_n^\pm| ,\qquad |\Phi_n^\pm\rangle \equiv \frac{1}{\sqrt{2}}\left(|++\rangle_n \pm i |--\rangle_n\right), \label{bell3}
\eeq
where the subscript $n$ means that the spin eigenstates are defined with respect to the $\hat{n}$ axis.

\subsection{Transverse photon}

We now turn to the transversely polarized case that has a much richer structure from the viewpoint of entanglement. The amplitude is proportional to \cite{Braun:2005rg} 
\beq
&& \epsilon^i\bar{u}_\alpha(\tilde{k})\Biggl[-m \gamma^i\gamma^+ I_L^g +  k_{\perp j}\gamma^j \gamma^i\gamma^+(2C_FI_T^{q_1}+\bar{z}I_T^g)+k_{\perp j}\gamma^i\gamma^j \gamma^+(2C_FI_T^{q_2}-zI_T^g)\Biggr]v_{\alpha'}(\tilde{k}') \nn
&&=
e^{\eta}\epsilon^i\bar{u}_\alpha(k)\Biggl[-m \gamma^i\gamma^+ I_L^g + k_\perp^i\gamma^+X +i\epsilon^{ij}k_{\perp j}\gamma^+\gamma_5Y \Biggr]v_{\alpha'}(k') , \label{trans}
\eeq
where we defined 
\beq
X=2C_F(I_T^{q_1}+I_T^{q_2})+(\bar{z}-z)I_T^g , \qquad Y=2C_F(I_T^{q_1}-I_T^{q_2})+I_T^g.  \label{AB}
\eeq  
We used the fact that $\gamma^{i=1,2}$ and $\gamma_5$ are invariant under the boost.  
Since we focus on the $\phi$-independent part of the cross section, we may set $\phi=0$ without losing generality.  Squaring (\ref{trans}) and summing over the polarizations of the incoming photon $\epsilon^i\epsilon^{j*}\to \delta^{ij}$,  we find that the cross section is proportional to
\beq
&& k_\perp^2|X|^2 \bar{u}_{\alpha} ( k )  \gamma^+  v_{\alpha'}( k')\bar{v}_{\beta'} ( k' )  \gamma^+  u_{\beta}( k)   +
k_\perp^2|Y|^2  \bar{u}_{\alpha} ( k )  \gamma^+\gamma_5  v_{\alpha'}( k')\bar{v}_{\beta'} ( k' )  \gamma^+\gamma_5  u_{\beta}( k)   \nn 
&&- m^2|I_L^g|^2   \bar{u}_{\alpha} ( k )  \gamma^+\gamma^i  v_{\alpha'}( k')\bar{v}_{\beta'} ( k' )  \gamma^+\gamma^i  u_{\beta}( k) \nn
&& - m k_{\perp} \bigg[ XI_L^{g*}\bar{u}_{\alpha} ( k )  \gamma^+  v_{\alpha'}( k') \bar{v}_{\beta'} ( k' )  \gamma^+ \gamma^1  u_{\beta}( k)  
+  X^*I_L^g \bar{u}_{\alpha} ( k )  \gamma^1 \gamma^+  v_{\alpha'}( k')\bar{v}_{\beta'} ( k' )  \gamma^+  u_{\beta}( k)    \bigg] \nn && - i m k_{\perp}   \bigg[ YI_L^{g*}\bar{u}_{\alpha} ( k )  \gamma^+ \gamma^5 v_{\alpha'}( k') \bar{v}_{\beta'} ( k' )  \gamma^+ \gamma^2  u_{\beta}( k)   +  Y^*I_L^g \bar{u}_{\alpha} ( k )   \gamma^+ \gamma^2 v_{\alpha'}( k')\bar{v}_{\beta'} ( k' )  \gamma^+\gamma^5  u_{\beta}( k)   \bigg] .
\eeq 
Similar to \cite{Fucilla:2025kit}, we evaluate the spinor products in the $(\hat{n},\hat{r},\hat{k})$ frame. The gamma matrices in this frame are related to those in the original frame as, at $\phi=0$, 
\beq
 \gamma^1= -\cos\theta \gamma^r + \sin \theta \gamma^k, \qquad 
\gamma^2=  \gamma^n.
\eeq
We then express the cross section in the form (\ref{lsigma}), (\ref{paramet}) with $L\to T$. 
The spin-averaged part is as in  (\ref{crossT}), 
\beq
A^T&=& \frac{\alpha_{em}e_q^2\alpha_s^2}{16\pi^2N_c(1-\xi^2)(k_\perp^2+\mu^2)^4}  \left( k_\perp^2(|X|^2+|Y|^2)+2m^2|I_L^g|^2\right).
\eeq
For the polarization vectors, 
we find $B^{T}_{r,k}=\bar{B}^{T}_{r,k}=0$,  but the $n$-component turns out to be  nonvanishing   
\beq
B_n^{T}
=\frac{-2mk_\perp {\rm Im}[(X+Y)I_L^{g*}]}{k_\perp^2(|X|^2+|Y|^2)+2m^2|I_L^g|^2},
\qquad 
 \bar{B}_n^{T}=
 \frac{-2mk_\perp {\rm Im}[(X-Y)I_L^{g*}]}{k_\perp^2(|X|^2+|Y|^2)+2m^2|I_L^g|^2}. \label{b}
\eeq
This means that the quark and the antiquark are transversely (i.e., normal to the production plane,  along the $\hat{n}$-direction in Fig.~\ref{kinematics}) polarized. It can be readily shown that
\beq
|B_n^T|,|\bar{B}_n^T|\le \frac{\left|{\rm Im}[(X\pm Y)I_L^{g*}]\right|}{\sqrt{2|I_L^{g*}|^2(|X|^2+|Y|^2)}} \le  \frac{\left|X\pm Y\right|}{\sqrt{2(|X|^2+|Y|^2)}}\le 1, \label{lessthan1}
\eeq
where in the first inequality we maximized in $k_\perp$. The last inequality is saturated when $X=Y$ and $X=-Y$, which occurs at $z=0$ and $\bar{z}=0$, respectively. 
Since the colliding particles $e+p$ or $\gamma+p$ are unpolarized, (\ref{b}) can be interpreted as  single spin asymmetry (SSA), analogous to the  transverse polarization of hyperons in unpolarized hadron collisions \cite{Bunce:1976yb,Lundberg:1989hw,HERMES:2007fpi,COMPASS:2021bws}.\footnote{The transverse polarization of final state particles can also arise if initial state particles are transversely polarized. See \cite{Cheng:2025zaw} for a recent study at the EIC.  } As is well known,  SSA requires a  phase, and therefore it cannot be generated in tree-level calculations in parity-conserving theories like QCD. Theoretical descriptions usually invoke higher-twist  distribution and/or fragmentation functions, or their transverse momentum dependent generalizations   \cite{Liang:1997rt,Kanazawa:2000cx,Anselmino:2001js,Zhou:2008fb,Koike:2017fxr,Kang:2021kpt}.  However, a phase can also be generated in perturbation theory beyond the tree level \cite{Dharmaratna:1996xd,Benic:2024fvk}, and this is what happens here.  Since the scattering amplitudes (\ref{I}) are obtained by convoluting GPDs with a hard factor, they are naturally complex.  
(Notice the $i\epsilon$ terms in (\ref{I}).)  The equation (\ref{b}) clearly shows that the nonvanishing polarization results from the interference between their real and imaginary parts. In contrast, in the Regge limit,  $X,Y$ and $I_L^g$ become purely imaginary, see the next subsection. This is why no polarization was found in  \cite{Fucilla:2025kit}. 
  
As for the $C^T$ matrix, we obtain
\beq
C^T_{nn}= \frac{(|X|^2-|Y|^2)k_\perp^2}{k_\perp^2(|X|^2+|Y|^2)+2m^2|I_L^g|^2},\label{cnn} 
\eeq
\beq
C^T_{rr}
&=&\frac{k_\perp^2}{k_\perp^2(|X|^2+|Y|^2)+2m^2|I_L^g|^2}\Biggl[-\frac{k_\perp^2-(1-2z)^2m^2}{k_\perp^2+(1-2z)^2m^2}|X|^2 +\frac{k_\perp^2-m^2}{k_\perp^2+m^2}|Y|^2 \\ && \qquad  + \frac{8z\bar{z}m^4|I_L^g|^2}{(k_\perp^2+(1-2z)^2m^2)(k_\perp^2+m^2)} + 4 m^2 \left( -  \frac{(1-2z){\rm Re}[X I_L^{g*}]}{k_\perp^2+(1-2z)^2m^2}  +\frac{{\rm Re}[Y I_L^{g*}] }{k_\perp^2+m^2}  \right) \Biggr] , \notag
\eeq
\beq
C^T_{kk} 
&=&\frac{1}{k_\perp^2(|X|^2+|Y|^2)+2m^2|I_L^g|^2}\Biggl[k_\perp^2\frac{k_\perp^2-(1-2z)^2m^2}{k_\perp^2+(1-2z)^2m^2} |X|^2+k_\perp^2\frac{k_\perp^2-m^2}{k_\perp^2+m^2}|Y|^2 \\ && \qquad  - \frac{2m^2(k_\perp^4-(1-2z)^2m^4)|I_L^g|^2}{(k_\perp^2+(1-2z)^2m^2)(k_\perp^2+m^2)} + 4 m^2k_\perp^2 \left(   \frac{(1-2z){\rm Re}[X I_L^{g*}]}{k_\perp^2+(1-2z)^2m^2}  +\frac{{\rm Re}[Y I_L^{g*}] }{k_\perp^2+m^2}  \right) \Biggr] ,\notag
\eeq
\beq
C^T_{rk}&=& 
\frac{2k_\perp m}{k_\perp^2(|X|^2+|Y|^2)+2m^2|I_L^g|^2} \Biggl[\frac{(1-2z)k_\perp^2|X|^2}{k_\perp^2+(1-2z)^2m^2} + \frac{k_\perp^2|Y|^2}{k_\perp^2+m^2} \\
&& \quad - \frac{2(1-z)m^2(k_\perp^2+(1-2z)m^2)|I_L^g|^2}{(k_\perp^2+m^2)(k_\perp^2+(1-2z)^2m^2)}  - \frac{k_\perp^2-(1-2z)^2m^2}{k_\perp^2+(1-2z)^2m^2}{\rm Re}[XI_L^{g*}] - \frac{k_\perp^2-m^2}{k_\perp^2+m^2}{\rm Re}[YI_L^{g*}]\Biggr], \notag \eeq
\beq
C^T_{kr}&=& 
\frac{2k_\perp m}{k_\perp^2(|X|^2+|Y|^2)+2m^2|I_L^g|^2} \Biggl[\frac{(1-2z)k_\perp^2|X|^2}{k_\perp^2+(1-2z)^2m^2} - \frac{k_\perp^2|Y|^2}{k_\perp^2+m^2} \label{ckr}\\
&& \qquad + \frac{2zm^2(k_\perp^2-(1-2z)m^2)|I_L^g|^2}{(k_\perp^2+m^2)(k_\perp^2+(1-2z)^2m^2)}  - \frac{k_\perp^2-(1-2z)^2m^2}{k_\perp^2+(1-2z)^2m^2}{\rm Re}[XI_L^{g*}] + \frac{k_\perp^2-m^2}{k_\perp^2+m^2}{\rm Re}[YI_L^{g*}]\Biggr]. \notag
\eeq
Unlike in the longitudinal case (\ref{cij}), the $C^T$ matrix  depends on the nonperturbative structure of the proton  through their dependence on  the quark and gluon GPDs. Importantly, the resulting  transverse density matrix $\rho^T$ in general represents a mixed state ${\rm Tr}[(\rho^T)^2]<1$ \cite{Fucilla:2025kit}. This is due to the combined effect of the transverse photon polarization sum and finite  longitudinal momentum transfer $2\xi P^-$  from the proton to the $q\bar{q}$ pair in the form of GPD convolutions (${\bf \Delta}\approx 0$ in the present approximation). On the other hand, the proton spin does not affect the entanglement property of the pair because we sum over the initial and final proton spins and there is no spin flip during scattering. The density matrix would take a  similar form if the target were a spinless pion, with the pion GPDs replacing the proton GPDs.

One might have expected that $B^T_n(z)=\bar{B}^T_n(z)$ and $C^T_{rk}(z)= C^T_{kr}(z)$ from CP symmetry, as in the case of $q\bar{q}\to t\bar{t}$ and $gg\to t\bar{t}$ productions   in QCD \cite{Bernreuther:2015yna}. However, neither of these relations hold in the present problem since the initial state $\gamma+{\mathbb P}$ is asymmetric (see also \cite{Qi:2025onf}). Instead,  the following relations hold exactly
\beq
B^T_n(z)=-\bar{B}^T_n(\bar{z}), \qquad C^T_{rk}(z)=-C^T_{kr}(\bar{z}).  \label{cp}
\eeq
(Note that $1-2z\to -(1-2z)$ under the transformation $z\to \bar{z}=1-z$.) 
For the valence quarks which we treat as massless $m_{u,d}=0$, this is trivial. For massive quarks $q=s,c,b$, (\ref{cp}) is a consequence of the property $F_{q}(x,\xi)=-F_q(-x,\xi)$ of the nonvalence quark GPDs which results in the relations 
\beq
X(z)=-X(\bar{z}), \qquad Y(z)=Y(\bar{z}).
\eeq
We may regard (\ref{cp}) as constraints from CP symmetry, since $z\to \bar{z}$ (or $\cos\theta\to -\cos\theta$ in the CM frame) interchanges the quark and the antiquark. 

\subsection{Regge limit}
Let us quickly check consistency with the result of \cite{Fucilla:2025kit} obtained in the Regge limit $\xi \to 0$. In this limit, the quark exchange contributions $I_L^q,I_T^{q_{1}},I_T^{q_2}$  are subdominant and neglected altogether. The gluon exchange contributions $I_L^g,I_T^g$ are dominantly imaginary, coming from the $i\epsilon$ prescription in the integrals  (\ref{I}). Keeping only the most singular term in the imaginary part, one finds  \cite{Braun:2005rg}
\beq
I_L^g
\approx   2\pi i \frac{ k_\perp^2-\mu^2}{k_\perp^2+\mu^2}   H_g(\xi,\xi), \qquad 
I_T^g 
\approx  2i\pi\frac{-2\mu^2}{k_\perp^2+\mu^2}  H_g(\xi,\xi),
\eeq
where it is assumed that the gluon GPD has a Regge behavior $H_g(\xi,\xi)\sim 1/\xi^\alpha$ with $\alpha\ll 1$. 
On the other hand, the result of \cite{Fucilla:2025kit} is expressed in terms of the following two integrals 
\beq
T_1= \int \frac{d^2{\bm p} T(p)}{({\bm p}-{\bm k})^2+\mu^2}, \qquad 
T_2= -\frac{1}{k_\perp^2}\int d^2p \frac{{\bm p}\cdot {\bm k} T(p)}{({\bm p}-{\bm k})^2+\mu^2},
\eeq
where $T$ is the T-matrix of the $q\bar{q}$ pair (`color dipole') scattering off the target proton. To make a connection with the collinear GPD calculation, we expand the integrand in powers of the intrinsic transverse momentum ${\bm p}$ and use the relation between $T(p)$ and $H_g(\xi,\xi)$   \cite{Hatta:2017cte}. We find 
\beq
T_1 &\approx& \frac{k_\perp^2-\mu^2}{(k_\perp^2+\mu^2)^3} \int d^2{\bm p}\, {\bm p}^2 T(p)  \approx \frac{k_\perp^2-\mu^2}{2(k_\perp^2+\mu^2)^3} \alpha_s  H_g(\xi,\xi),
\nn
T_2 &\approx& -\frac{1}{(k_\perp^2+\mu^2)^2}\int d^2{\bm p}\, {\bm p}^2 T(p) \approx -\frac{1}{2(k_\perp^2+\mu^2)^2}\alpha_s  H_g(\xi,\xi).
\eeq
This gives the correspondence 
\beq
T_1 \approx \frac{\alpha_sI_L^g}{4\pi i (k_\perp^2+\mu^2)^2} , \qquad T_1+T_2\approx \frac{\alpha_sI_T^g}{4\pi i (k_\perp^2+\mu^2)^2} . \label{zer}
\eeq
Inserting these relations into $A^{T/L}$,  and noting that $X\approx (\bar{z}-z)I_T^g$, $Y\approx I_T^g$, one recovers the dijet cross section originally calculated in the $k_\perp$-factorization framework  \cite{Nikolaev:1994cd,Bartels:1996ne}, as already noted in \cite{Braun:2005rg}. Furthermore, inserting  into  (\ref{cnn})-(\ref{ckr}), we find full agreement with \cite{Fucilla:2025kit}. Note that  $T_1$ linearly vanishes as $k_\perp \to \mu$ \cite{Fucilla:2025kit}. Consistently, $I_L^g$ has the same behavior in the present approximation. 

\section{Entanglement, Bell nonlocality and magic}

In this section, we discuss three different measures of spin-spin correlation between the produced  quark and  antiquark: entanglement,   Bell nonlocality and magic. The spin density matrix formalism allows one to apply these key concepts in quantum information science to relativistic field theories.  Numerical results will be presented in the next section. 

\subsection{Entanglement}
According to the Peres-Hordecki criterion \cite{Peres:1996dw,Horodecki:1997vt}, for generic two-qubit systems,  the necessary and sufficient condition for separability (i.e., no entanglement) is that the partial transpose of the density matrix 
\beq
\rho^{\cal T}\equiv \frac{1}{4}\left( {\mathds 1}\otimes {\mathds 1}^{\cal T} +B_a \sigma^a\otimes {\mathds 1}^{ \cal T}+\bar{B}_b {\mathds 1}\otimes  (\sigma^b)^{\cal T}+ C_{ab}\sigma^a \otimes (\sigma^b)^{\cal T} \right), 
\eeq
is nonnegative definite. (The symbol ${\cal T}$ denotes `transpose.') In the present problem with $\rho=\rho^{L/T}$, it is convenient to  formally introduce the spin eigenstates  $|\pm\rangle_n=\frac{1}{\sqrt{2}}(|+\rangle_k\pm |-\rangle_k)$ along the direction $\hat{n}$. In the basis $\{|++\rangle_n,|+-\rangle_n,|-+\rangle_n,|--\rangle_n\}$,  
the (transposed) density matrix is block diagonal \cite{Afik:2020onf}
\beq
&&\hspace{-3mm}\rho^{\cal T}= \label{basis} \\
&& \hspace{-3mm}\frac{1}{4}\begin{pmatrix} 1+B_n+\bar{B}_n+C_{nn} & 0 & 0 & C_{rr}+C_{kk}+i(C_{rk}-C_{kr}) \\ 
0 & 1+B_n-\bar{B}_n -C_{nn} & C_{rr}-C_{kk}-i(C_{rk}+C_{kr}) & 0 \\ 0 & C_{rr}-C_{kk}+i(C_{rk}+C_{kr}) & 1-B_n+\bar{B}_n-C_{nn} & 0 \\ 
C_{rr}+C_{kk}-i(C_{rk}-C_{kr}) & 0 & 0 & 1-B_n-\bar{B}_n+C_{nn} \end{pmatrix}. \notag
\eeq
In order for this matrix to be nonnegative, the following conditions need to be satisfied 
\beq
1\pm C_{nn}\ge 0, \qquad 
(1\pm C_{nn})^2-(B_n\pm \bar{B}_n)^2-(C_{rr}\pm C_{kk})^2-(C_{rk}\mp C_{kr})^2 \ge 0.
\eeq
Thus the necessary and sufficient condition for entanglement is 
\beq
{\rm max}\{\Delta_1,\Delta_2, |C_{nn}|-1\}>0, \label{max} 
\eeq
where 
\beq
&&\Delta_1\equiv  \sqrt{(C_{rr}-C_{kk})^2+(C_{rk}+C_{kr})^2+(B_{n}-\bar{B}_{n})^2}-|1-C_{nn}|, \nn  
&& \Delta_2\equiv  \sqrt{(C_{rr}+C_{kk})^2+(C_{rk}-C_{kr})^2+(B_{n}+\bar{B}_{n})^2}-|1+C_{nn}|.  \label{delta12}
\eeq
 Note that  nonnegativity of the density matrix ensured by the condition (\ref{max}) is not affected by unitary Wigner rotations. This is the physical reason behind the general statement that entanglement is frame-independent \cite{Peres:2002wx}.   
In \cite{Fucilla:2025kit}, it has been noticed that in the Regge limit,  
\beq
\Delta^{L/T}_2=-\Delta^{L/T}_1 = -2C^{L/T}_{nn}. \label{rem}
\eeq
In the longitudinal case, the $q\bar{q}$ pair is always maximally  entangled because  $C_{nn}^L=1$ and $\Delta_1^L=2>0$. In the transverse case where $\Delta_2^T=-2C_{nn}^T\ge 0$, the pair is almost always entangled except when $C_{nn}^T=0$, which occurs in  certain kinematical limits such as $k_\perp=0$ or $k_\perp=\infty$.  
The equation (\ref{rem}) is a direct consequence of the following relations obeyed by the components of $C^{L/T}$ 
\beq
(C_{rr})^2+(C_{rk})^2+(C_{kr})^2+(C_{kk})^2 -(C_{nn})^2 -1= 0, \label{id1}
\qquad C_{nn} +C_{rr}C_{kk}-C_{rk}C_{kr}=0. 
\eeq
Away from the Regge limit, (\ref{id1}) continue to hold in the longitudinal case. However, in the transverse case we find  from (\ref{cnn})-(\ref{ckr})
\beq
\begin{split}
&(C_{rr}^T)^2+(C_{rk}^T)^2+(C_{kr}^T)^2+(C_{kk}^T)^2 -(C_{nn}^T)^2 -1 \\  
& \qquad  = 8k_\perp^2m^2\frac{ ({\rm Re}[XI_L^{g*}])^2+({\rm Re}[YI_L^{g*}])^2-(|X|^2+|Y|^2)|I_L^g|^2}{((|X|^2+|Y|^2)k_\perp^2+2|I_L^g|^2m^2)^2}, \label{id3}
\\
&C_{nn}^T +C_{rr}^TC_{kk}^T-C_{rk}^TC_{kr}^T = 4k_\perp^2m^2\frac{ -({\rm Re}[XI_L^{g*}])^2+({\rm Re}[YI_L^{g*}])^2+(|X|^2-|Y|^2)|I_L^g|^2}{((|X|^2+|Y|^2)k_\perp^2+2|I_L^g|^2m^2)^2} . 
\end{split}
\eeq
In the Regge limit, the right hand sides vanish because $X,Y,I_L^g$ are purely imaginary. However, in general $X,Y,I_L^g$ have both real and imaginary parts.\footnote{ 
The right hand sides also vanish if $X,Y,I_L^g$ are all relatively real. Again, this not the case in general. }  Therefore, in the present calculation  (\ref{rem}) does not hold, and the condition (\ref{max}) becomes nontrivial to satisfy. We expect to find regions in phase space where the $q\bar{q}$ pair is separable, i.e., not entangled. Interestingly,  the right hand sides of (\ref{id3}) vanish for massless quarks $m=0$. This means that the $u\bar{u}$ and $d\bar{d}$ pairs, and also the $s\bar{s}$ pair in practice, are always entangled.  

Let us elaborate more on the massless case. The $C^T$ matrix becomes diagonal 
\beq
C^T_{ab}=\begin{pmatrix} \frac{|X|^2-|Y|^2}{|X|^2+|Y|^2} & 0 & 0 \\ 0 & -\frac{|X|^2-|Y|^2}{|X|^2+|Y|^2} & 0 \\ 0 & 0 & 1
\end{pmatrix}.
\eeq
The corresponding density matrix can be written in the form 
\beq
\rho^T= \frac{|X|^2}{|X|^2+|Y|^2} |\Phi^+\rangle\langle \Phi^+| + \frac{|Y|^2}{|X|^2+|Y|^2}|\Phi^-\rangle\langle \Phi^-|.
\label{rhot}
\eeq
This represents a mixed state between two  of the Bell states. 
\beq
|\Phi^\pm\rangle = \frac{1}{\sqrt{2}}\bigl( |++\rangle \pm |--\rangle\bigr). \label{phiminus}
\eeq
Eq.~(\ref{rhot}) is an example of the so-called `Bell diagonal' states \cite{Lang:2010xtw} which are mixed states comprised only of the Bell states 
\beq
\rho=p_1 |\Psi^+\rangle\langle \Psi^+| + p_2|\Psi^-\rangle \langle \Psi^-| + p_3|\Phi^+\rangle\langle \Phi^+| +p_4|\Phi^-\rangle\langle \Phi^-|, \label{massless}
\eeq
with $\sum_i p_i=1$. 
It is known that a Bell diagonal state is  separable if $p_i\le \frac{1}{2}$ for all $i$. In the present problem, the pair is separable only if $|X|=|Y|$ where  $C_{nn}^T=0$. If either $|X|$ or $|Y|$ (accidentally) vanishes,  then one of the Bell states is realized.
Note that in the Regge limit, $X\approx (1-2z)I_T^g$ and $Y\approx I_T^g$. Thus the state is separable when $z=0,1$, and  becomes the  Bell state $|\Phi^-\rangle$ at $z=\frac{1}{2}$.  Note that under Lorentz transformations, Bell states with sharp momenta remain Bell states up to local unitary transformations. 

\subsection{Bell nonlocality}

In terms of the correlation matrix $C$, the Bell-CHSH inequality reads  \cite{Bell:1964kc,Clauser:1969ny} 
\beq
{\rm Max}_{\{\vec{n}_i\}}  \Bigl| n_1^a C_{ab} (n_2^b+n_4^b) + n_3^aC_{ab}(n_2^b-n_4^b)
\Bigr| \le 2, \label{ch}
\eeq
where $\vec{n}_i$ ($i=1,2,3,4$) are unit vectors $|\vec{n}_{i}|=1$. Depending on the matrix $C_{ab}$, quantum two-qubit states can violate this inequality for deliberate choices of $\{\vec{n}_i\}$. When this happens, we say the $q\bar{q}$ pair exhibits  `Bell nonlocality,'  and this property is frame-independent  (see \cite{Barr:2024djo} and references therein).  
The necessary and sufficient condition for Bell nonlocality  is that the largest two of the three eigenvalues $\mu_3\le \mu_2\le \mu_1$ of the matrix $C^{\cal T}C$  satisfy \cite{Horodecki:1995nsk}
\beq
1<\mu_1+\mu_2\le 2. \label{eigen}
\eeq
For our problem, the three eigenvalues of 
\beq
C^{\cal T} C = \begin{pmatrix} C_{nn}^2 & 0 & 0 \\ 0 & C_{rr}^2+C_{kr}^2 & C_{rr}C_{rk}+C_{kr}C_{kk} \\
0 & C_{rr}C_{rk}+C_{kr}C_{kk} & C_{kk}^2+C_{rk}^2 \end{pmatrix},
\eeq
are $C_{nn}^2$ and 
\beq
 \frac{C_{rr}^2+C_{kr}^2+
C_{kk}^2+C_{rk}^2  \pm \sqrt{\left(C_{rr}^2+C_{kr}^2-
C_{kk}^2-C_{rk}^2\right)^2+4\left(C_{rr}C_{rk}+C_{kr}C_{kk}\right)^2}}{2}. \label{eigenvalues} 
\eeq
In the Regge limit, the identities (\ref{id1}) hold, and the above three eigenvalues reduce to 
\beq
\mu_1= 1, \qquad \mu_2=\mu_3 = C_{nn}^2.
\eeq
Thus the inequality is always maximally violated in the longitudinal case $C_{nn}^L=1$, while in the transverse case it is violated if $C_{nn}^T \neq 0$. Since $C_{nn}^T\le 0$, this happens to be the same condition as (\ref{max}), (\ref{rem}) for nonvanishing entanglement \cite{Fucilla:2025kit}. In other words,  Bell nonlocality and entanglement are equivalent. This is  remarkable because, in general, the states that exhibit Bell nonlocality form a subset of the states that are entangled \cite{Werner:1989zz}.   Away from the Regge limit,  the Bell-CHSH inequality continues to be maximally violated in the longitudinal case. In the transverse case, it is always violated for the $q=u,d$ quarks because  (\ref{id1}) also holds for massless quarks. For massive quarks,  there will be  kinematical regions  where the inequality is not violated. On general grounds, we expect that this region is broader than the region where the pair is separable (not entangled).

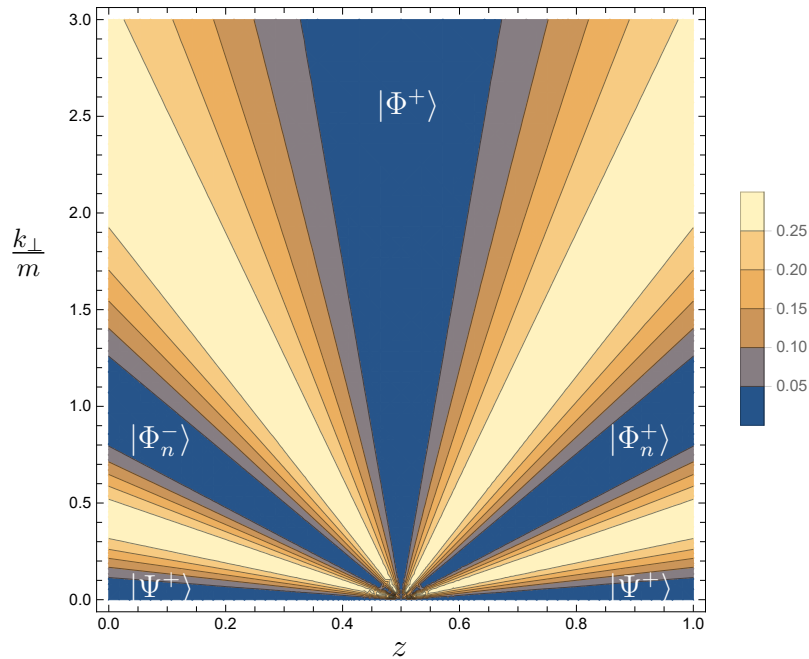
\begin{figure}[t]
    \centering
 \begin{overpic}[
         width=0.55\textwidth
         ]{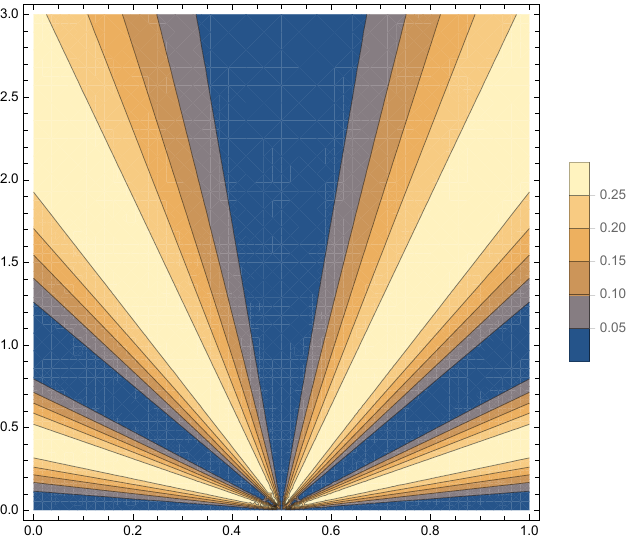}
           \put(-8,50){{\Large  $\frac{k_\perp}{m}$}}
            \put(43,-3){{\large $z$}}
              \put(41,70){{\color{white}{\large $|\Phi^+\rangle$}}}
              \put(72,5){{\large {\color{white}$|\Psi^+\rangle$}}}
               \put(8,5){{\large {\color{white}$|\Psi^+\rangle$}}}
                \put(72,25){{\large {\color{white}$|\Phi_n^+\rangle$}}}
              \put(8,25)  {{\large {\color{white}$|\Phi_n^-\rangle$}}}
         \end{overpic}
         \vspace{3mm}
    \caption{Stabilizer R\'enyi entropy $M_2^L$ (\ref{longre}) of a massive quark-antiquark pair produced by a longitudinally polarized virtual photon. $M_2^L$ is suppressed in the blue regions where the pair forms the Bell states (\ref{bell1}), (\ref{bell2}), (\ref{bell3}). }
    \label{magicL}
\end{figure}

\subsection{Magic}

Finally, let us consider another measure of entanglement called  `magic'. (See \cite{White:2024nuc} for an introduction friendly to high energy physicists.) In quantum computing, it is known that entanglement alone does not guarantee quantum advantage \cite{Gottesman:1998hu}. Entangled pairs with nonvanishing magic  are crucial for   quantum computers to outperform  classical ones.  Recently, this notion has been imported to the high energy physics community \cite{White:2024nuc,Liu:2025qfl,Gargalionis:2025iqs}. One may ask, for example, how much magic is carried by elementary particle pairs produced  in collider experiments. Of course, at present such a question is mostly of conceptual interest.  But it may offer new  perspectives on the nature of scattering in quantum field theories. 
To quantify magic, we adopt the `stabilizer R\'enyi entropy' \cite{Leone:2021rzd,White:2024nuc} which  takes the following form for the present problem and  in the present frame
\beq
M_2 = -\ln\left( \frac{1+B_n^4+\bar{B}_n^4+ C_{nn}^4+C_{rr}^4+C_{kk}^4+C_{kr}^4+C_{rk}^4}{1+B_n^2+\bar{B}_n^2+C_{nn}^2+C_{rr}^2+C_{kk}^2+C_{kr}^2+C_{rk}^2} \right). \label{stabi}
\eeq
Note that this quantity depends on the choice of frame and spinor basis \cite{Liu:2025qfl}. The following results should be understood as such.  

Let us study the longitudinal case which allows for a simple analytical treatment, deferring the transverse case to the numerical section. From (\ref{cij}), we find   a compact analytical  formula  
\beq
M_2^L=\ln \frac{(k_\perp^2+m^2(1-2z)^2)^4}{k_\perp^8+14k_\perp^4m^4(1-2z)^4+m^8(1-2z)^8}. \label{longre}
\eeq
This is plotted in Fig.~\ref{magicL}. By construction, $M_2$ vanishes for the Bell states $|\Phi^+\rangle$ (\ref{bell1}),  $|\Psi^+\rangle$ (\ref{bell2}), $|\Phi_n^\pm\rangle$ (\ref{bell3}) realized along the lines  $z=\frac{1}{2}$, $k_\perp=0$ and $\frac{k_\perp}{m}=|2z-1|$, respectively. Everywhere else, magic is nonzero even though the pair is always maximally entangled. $M_2^L$ takes the maximal value 
\beq
M_2^{L,{\rm max}}= \ln \frac{4}{3}\approx 0.288, \label{l2max}
\eeq
 along the following  four lines in the $(z,k_\perp)$ plane 
\beq
\frac{k_\perp}{m}=(\sqrt{2}\pm 1)|2z-1|, \label{maxline}
\eeq
which are clearly visible in Fig.~\ref{magicL}. Along these lines, 
the density matrix takes the form 
\beq
\rho^L=\frac{1}{2}\begin{pmatrix} 1 & 0 & 0 & \pm  \frac{1\pm i}{\sqrt{2}} \\ 0 & 0 & 0 & 0 \\ 0 & 0 & 0 & 0 \\ 
\pm \frac{ 1\mp i}{\sqrt{2}} & 0 & 0 & 1\end{pmatrix}, \label{maxrho}
\eeq
using the same basis as in (\ref{basis}). (The sign combination is arbitrary, except that the 14 and 41 components are complex conjugate to each other.) 
The result (\ref{l2max}) is significantly smaller than the theoretical upper bound for pure two-qubit states \cite{Liu:2025frx}
\beq
M_2^{\rm max}= \ln\frac{16}{7}\approx 0.827. \label{th}
\eeq
However, we point out that this limit can be saturated only by  states with nonzero polarization vectors, namely, $B,\bar{B}\neq 0$ in the generic parametrization (\ref{paramet}). Since  $B^L=\bar{B}^L= 0$, the maximal value (\ref{th}) cannot be expected. A better comparison may be the range of $M_2$ for the subset of states with vanishing polarization.  In Appendix A, we revisit  the derivation of \cite{Liu:2025frx} by imposing an extra constraint $B=\bar{B}=0$ and obtain a stronger bound 
\beq
\left. M_2^{\rm max}\right|_{B=\bar{B}=0}= \ln \frac{9}{5}\approx 0.588,
\label{minimum}
\eeq
which is still larger than (\ref{l2max}) by a factor of about 2. In Appendix A, we further show that (\ref{l2max}) is the maximum value of $M_2$ if one restricts to an even smaller subset of states. 
We note that the numbers $\ln \frac{9}{5}$ and $\ln\frac{4}{3}$ have been previously encountered in  \cite{Liu:2025qfl} as the maximal amount of magic that can be generated from certain initial stabilizer states in QED processes.

\section{Numerical results}

In this section we delineate  kinematical regions in the parameter space $(z,k_\perp)$ where the $q\bar{q}$ pair exhibits entanglement and Bell nonlocality. We also compute the stabilizer R\'enyi entropy $M_2^T$ for the transversely polarized photon.  We employ the Goloskokov-Kroll model \cite{Goloskokov:2006hr,Goloskokov:2007nt} of the quark and gluon GPDs.  This model has been fitted to reproduce the exclusive vector meson production data at HERA. At small skewness, the gluon GPD features the Regge behavior 
\beq
H_g(\xi,\xi,\mu_h)\sim \frac{1}{\xi^{0.1+0.06N}}, \qquad N=\ln \frac{\mu_h^2}{Q_0^2}, \label{l}
\eeq
with $Q_0^2=4\, {\rm GeV}^2$ and $\mu_h=Q$ in electroproduction \cite{Goloskokov:2006hr,Goloskokov:2007nt}).    Thus the model should produce results consistent with \cite{Fucilla:2025kit} in the Regge limit. Adapting to our process, we use $\mu_h= {\rm max}\{Q,m,k_\perp\}$.  The quark masses are fixed as  $m_u=m_d=0$, $m_s=0.093$ GeV, $m_c=1.27$ GeV and $m_b=4.18$ GeV. For light quark production $q=u,d,s$, we include both the quark and gluon GPD contributions.  For heavy quark production $q=c,b$, we include only the gluon GPD  contribution  (i.e., $I^q_{L/T}=0$ in this case) since the $c,b$-quark GPDs are unavailable (and expected to be negligible). 
Using the GK model, we numerically evaluate the integrals (\ref{I}) on  grid points on a $50\times 50$ lattice in the $(z,k_\perp)$ plane 
\beq
0.1\le z \le 0.9, \qquad 1\, {\rm GeV}\le k_\perp \le k_\perp^{\rm max},
\eeq
and smoothly interpolate the results.   $k_\perp^{\rm max}$ is chosen appropriately for a given value of the $\gamma +p$ CM energy $W$. The integrals can be evaluated in a numerically stable way. For very small values of $\xi$, this is achieved by using the contour deformation technique as explained around (43) of \cite{Hatta:2025vhs}.

In electroproduction at $Q\neq 0$, the measured cross section (\ref{ep}) is the linear combination of the transverse and longitudinal contributions.  
Unfortunately,  it is difficult to experimentally separate the two contributions at the EIC since typically $\varepsilon \approx 1$ at high energy colliders (see however \cite{Klest:2025yik}). This means that the density matrices must be averaged  
\beq
A^T\rho^T+\varepsilon A^L \rho^L=\frac{A^T+\varepsilon A^L}{4} \left({\mathds 1}\otimes {\mathds 1}+ B_n^{\rm DIS}\sigma^n \otimes {\mathds 1} + \bar{B}_n^{\rm DIS}{\mathds 1}\otimes \sigma^n+ C_{ab}^{\rm DIS}\sigma^a\otimes \sigma^b \right), 
\eeq
where 
\beq
B_n^{\rm DIS}= \frac{A^TB_n^T}{A_T+\varepsilon A^L}, \qquad \bar{B}_n^{\rm DIS}= \frac{A^T\bar{B}_n^T}{A_T+\varepsilon A^L}, \qquad C_{ab}^{\rm DIS}=\frac{A^TC_{ab}^T+\varepsilon A^LC_{ab}^L}{A^T+\varepsilon A^L}. \label{disbb}
\eeq
$\Delta_{1,2}$, $\mu_{1,2}$ and $M_2$ defined in the previous section  are calculated from these coefficients in the same way. In the following numerical results, we set $\varepsilon=1$.

\subsection{Polarization}

\begin{figure}[t]
    \centering

    \begin{minipage}{0.35\textwidth}
    \begin{overpic}[width=\textwidth]{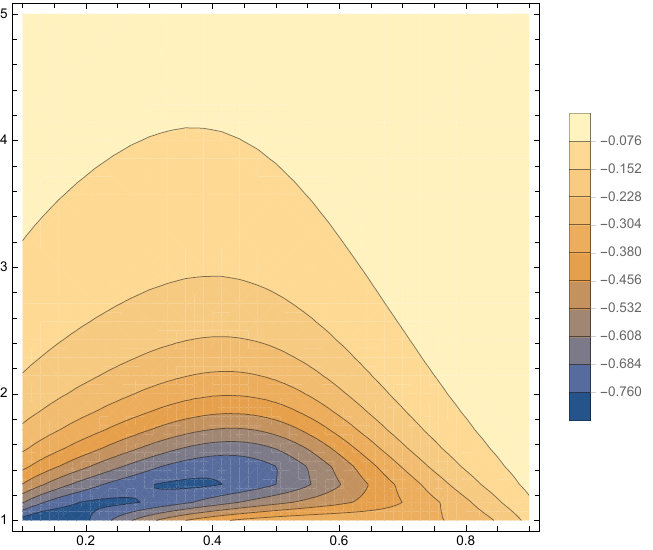}
    \put(75,78){{\large $s$}}
            \put(42,-5){$z$}
            \put(-10,40){$k_\perp$}
            \end{overpic}
    \end{minipage}
    \begin{minipage}{0.35\textwidth}
      \begin{overpic}[width=\textwidth]{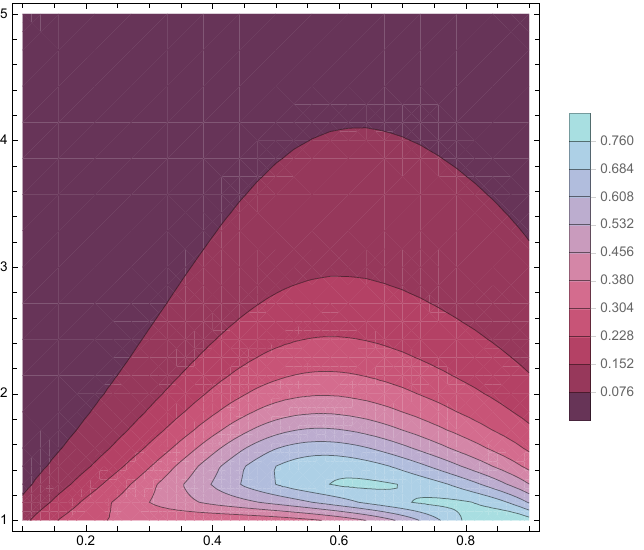}
      \put(8,78){{\color{white}{\large $\bar{s}$}}}
            \put(42,-5){$z$}
            \end{overpic}
    \end{minipage}
    \vspace{0.02\textwidth}
\begin{minipage}{0.35\textwidth}
    \begin{overpic}[width=\textwidth]{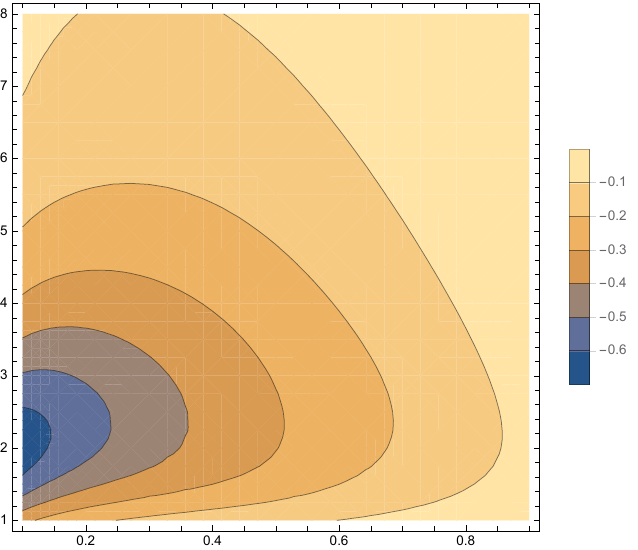}
    \put(75,79){{\large $c$}}
            \put(42,-5){$z$}
            \put(-10,40){$k_\perp$}
            \end{overpic}
    \end{minipage}
    \begin{minipage}{0.35\textwidth}
      \begin{overpic}[width=\textwidth]{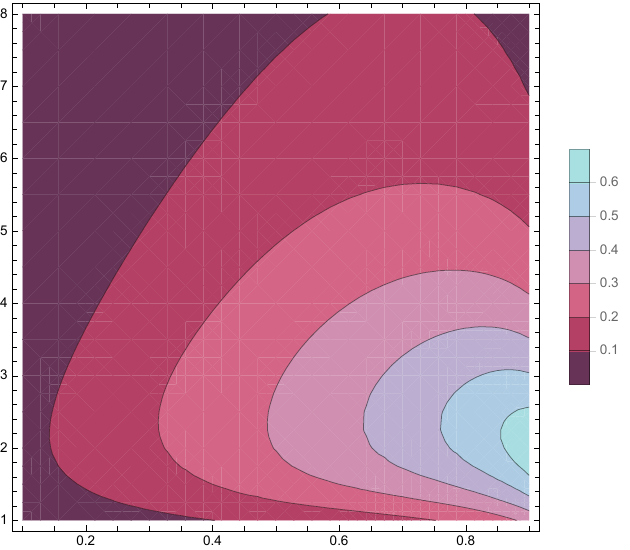}
      \put(8,79){{\color{white}{\large $\bar{c}$}}}
            \put(42,-5){$z$}
            \end{overpic}
    \end{minipage}

    \caption[*]{Upper panels: Polarization of the strange (left) and antistrange  (right) quarks  in UPC at $W=30$ GeV, $Q=0$ GeV. Lower panels: Polarization of the charm (left) and anticharm (right) quarks in the same process. Here and in all the plots below, $k_\perp$ in the vertical axis is in units of GeV. }
    \label{bn}
\end{figure}
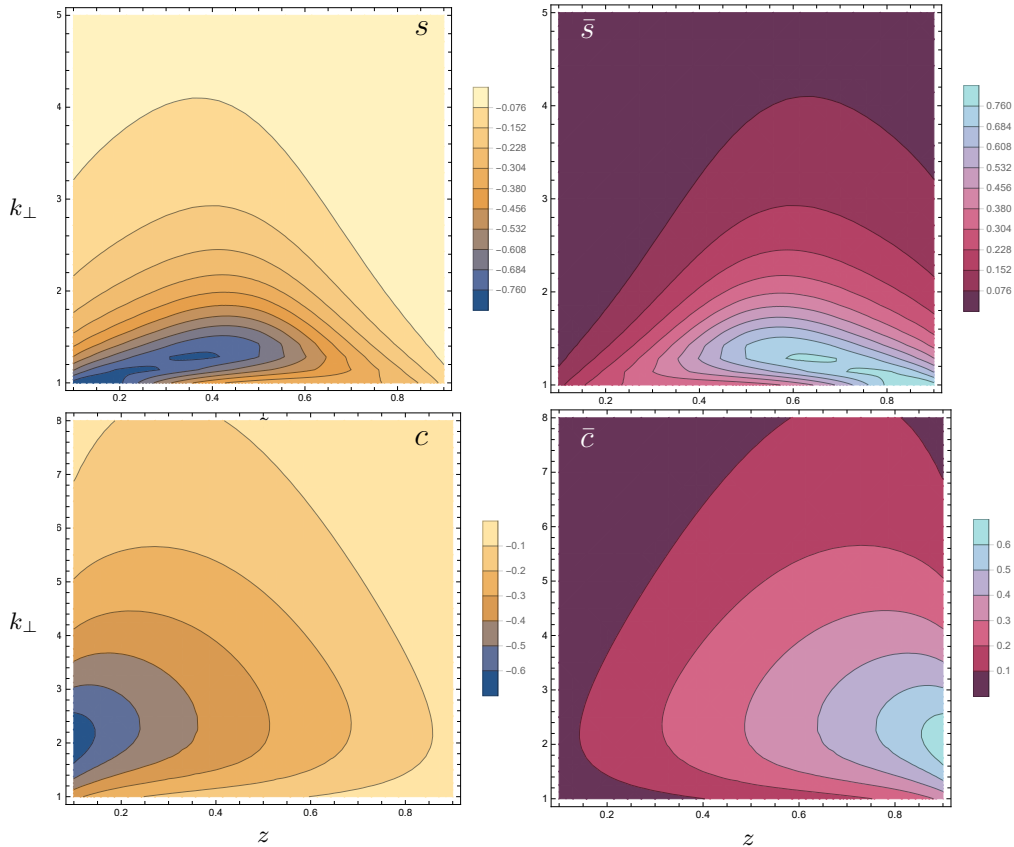

First we show the results on the polarization  of the quark $B_n$ and the antiquark $\bar{B}_n$. As we noted already,  $B_n$ and $\bar{B}_n$ are nonvanishing only for massive quarks and for the transversely polarized photon, and tend to be suppressed at high CM energies. Therefore, it is best to focus on low-$W$  UPC or low-$W$ electroproduction at low-$Q^2$.  In Fig.~\ref{bn}, we plot $B_n$ and $\bar{B}_n$ for the strange (upper panels) and charm (lower panels) quarks in UPC at $W=30$ GeV. We find  that $B_n$ is negative and $\bar{B}_n$ is positive for all massive quark flavors $q=s,c,b$, consistently with (\ref{cp}). The net polarization along the $\hat{n}$ direction $B_n(z)+\bar{B}_n(z)\propto z-\bar{z}$    vanishes at $z=\frac{1}{2}$. 
 The peak structure in $k_\perp$ is as expected from the analytic formula (\ref{b}), with the peak magnitude reaching 80\%, 60\%, 50\% for strange, charm and  bottom quarks, respectively.\footnote{It should be mentioned that $B_n$ of the strange quark is largest in the low-momentum region $k_\perp \gtrsim 1$ GeV where, strictly speaking,  the present perturbative approach is not fully justified in photoproduction $Q=0$.}   These peaks occur when the quark (antiquark) is `soft' $z\ll 1$ ($\bar{z}\ll 1$)  in the original frame (\ref{fast}) in which the photon is fast moving and the $q\bar{q}$ pair is produced in the forward rapidity region, cf., the comment below (\ref{lessthan1}).  Negative $B_n$ means that the quark is polarized in the direction of $-\vec{q}\times \vec{k}_q$, see Fig.~\ref{kinematics}. Similarly, the antiquark is  polarized along the direction $-\vec{q}\times \vec{k}_{\bar{q}}$.

Comments are in order regarding the magnitude of  the polarization.  In  semi-inclusive quark or antiquark production, the imaginary part of the amplitude responsible for nonvanishing polarization is suppressed by a factor of $\alpha_s$ since it requires a loop. Typically, one finds (sub-)percent level asymmetries in such perturbative calculations  \cite{Dharmaratna:1996xd,Benic:2024fvk}. However, in exclusive production with  color singlet exchanges in the $t$-channel, the real and imaginary parts of the scattering amplitudes are of the same order in $\alpha_s$. Consequently, $B_n,\bar{B}_n$ (\ref{b})  are order unity instead of order $\alpha_s$. Still, a polarization as large as 50-80\% is quite remarkable. To our knowledge, such a strong polarization of perturbative origin has not been reported in the QCD literature. 

It is generally expected that the quark-level polarization is largely retained during fragmentation into heavy baryons $q\to \Lambda_q$, $\bar{q}\to \bar{\Lambda}_q$  \cite{Galanti:2015pqa}. In particular, the process $s\to \Lambda\to p+\pi^-$ can be used as a polarimeter of the strange quark.  
The HERMES collaboration \cite{HERMES:2007fpi} measured the polarization of $\Lambda$ and $\bar{\Lambda}$ in semi-inclusive DIS in the photoproduction region $Q\approx 0$. They found a positive polarization for $\Lambda$ but almost zero  polarization for $\bar{\Lambda}$. On the other hand, the COMPASS collaboration \cite{COMPASS:2021bws}  found very small values consistent with zero for both $\Lambda$ and $\bar{\Lambda}$. The contribution from exclusive $s\bar{s}$ production, if any at all in these low energy experiments, will be significantly diluted by the inclusive background. On the other hand, diffractive events with a rapidity gap are copious at the EIC. By triggering on such events, one should  be able to test our prediction in future.

\subsection{Entanglement}

\begin{figure}[t]
    \centering
\hspace{5mm}
    \begin{minipage}{0.3\textwidth}
    \begin{overpic}[width=\textwidth]{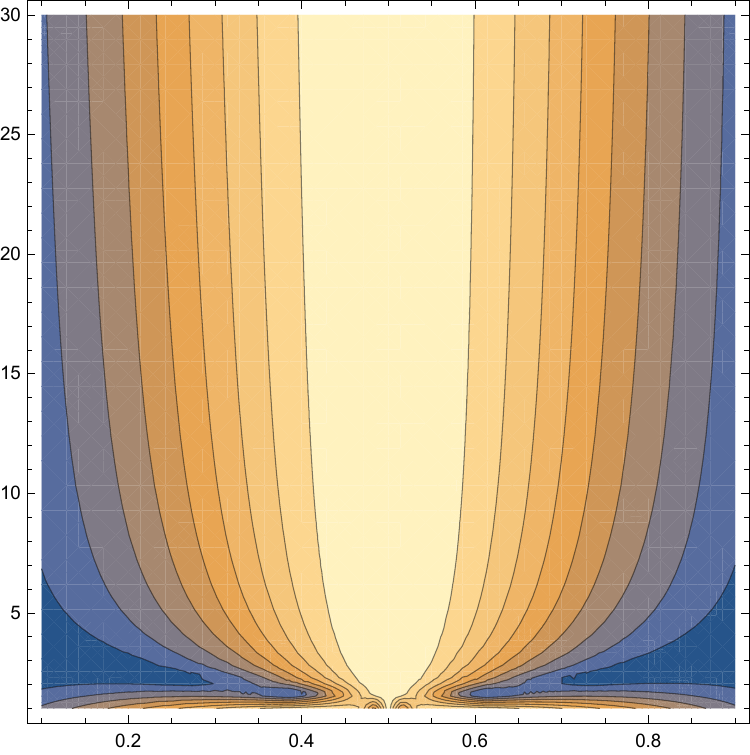}
    \put(8,90){\colorbox{white}{\textbf{$u\bar{u}$}}}
            \put(50,-7){$z$}
             \put(44,60){ $|\Phi^-\rangle$}
            \put(-10,50){$k_\perp$}
            \end{overpic}
    \end{minipage}
    \begin{minipage}{0.3\textwidth}
      \begin{overpic}[width=\textwidth]{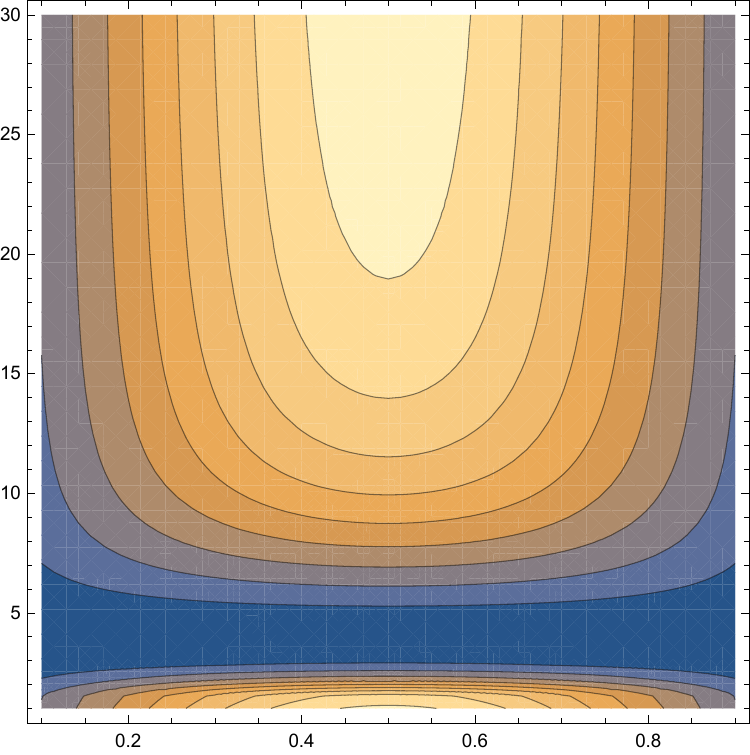}
      \put(8,90){\colorbox{white}{\textbf{$c\bar{c}$}}}
            \put(50,-7){$z$}
              \put(44,85){ $|\Phi^-\rangle$}
            \end{overpic}
    \end{minipage}
 \begin{minipage}{0.348\textwidth}
 \vspace{-1mm}
      \begin{overpic}[width=\textwidth]{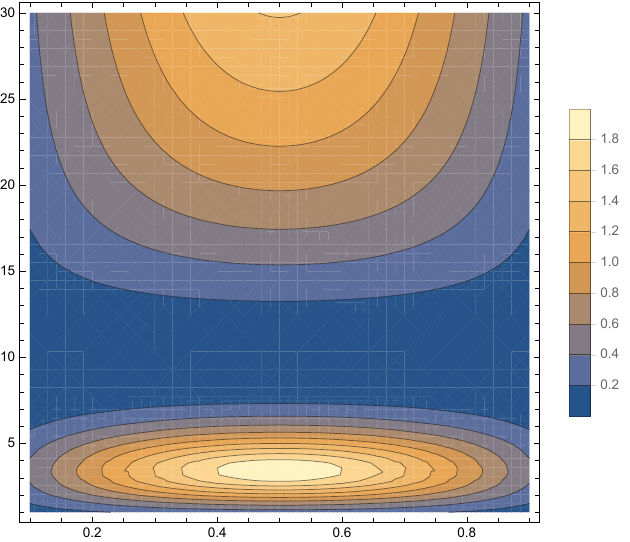}
       \put(8,76){\colorbox{white}{\textbf{$b\bar{b}$}}}
           \put(47,-7){$z$}
           \put(37,10){  $|\Psi_n^-\rangle$}
            \end{overpic}
    \end{minipage}
    \vspace{0.02\textwidth}

    \caption[*]{Left hand side of (\ref{max})  in UPC at $W=500$ GeV, $Q=0$. Left: $u\bar{u}$ pair, Middle: $c\bar{c}$ pair, Right: $b\bar{b}$ pair.  Entanglement is stronger in  brighter regions where mostly ${\rm max} \{\Delta_1,\Delta_2,|C_{nn}-1\}=\Delta_2$. The pair reaches  maximally entangled states   $\Phi^-$ (\ref{phiminus}) and $\Psi_n^-$ (\ref{psiminus}) with $\Delta_2\approx 2$ in the indicated regions. }
    \label{1}
\end{figure}

The degree of entanglement between the quark and antiquark spins can be quantified by $\Delta_{1,2}$ introduced in (\ref{delta12}).  First, consider UPCs at the LHC  and fix $W=500$ GeV, $Q=0$ and $k_\perp^{\rm max}=30$ GeV.   In Fig.~\ref{1}, we plot the left hand side of (\ref{max}) in the $(z,k_\perp)$ plane for $u\bar{u}$ (left), $c\bar{c}$ (middle) and  $b\bar{b}$ (right) pairs, respectively.  The results for $d\bar{d}$, $s\bar{s}$ pairs are similar to the $u\bar{u}$ case. In practice, these are plots of $\Delta_2^T$ that dominates in (\ref{max}).  Since $W$ is typically quite large at the LHC, the results are close to those in the Regge limit  calculation in  \cite{Fucilla:2025kit}. In particular, the pairs are everywhere entangled (i.e., (\ref{max}) is always satisfied).  Light quarks are maximally entangled ($\Delta^T_2\approx 2$) along the line $z=\frac{1}{2}$ as discussed around (\ref{massless}). For massive  quarks, an almost maximally entangled state
\beq
|\Psi^-_n\rangle \approx \frac{1}{\sqrt{2}}\left(|+-\rangle_n-i|-+\rangle_n\right), \label{psiminus}
\eeq
is realized at a peak around $z=\frac{1}{2}$ and $k_\perp \approx m$. The peak occurs because $I_g^L\approx 0$ (see the comment below (\ref{zer})) and hence $C_{nn}^T\approx -1$ in this region. The peak is followed by  a dip as $k_\perp$  increases   (i.e., entanglement becomes weak) before rising again at higher $k_\perp$ where the mass effect becomes less important.   It should be noted  that, when  $W=500$ GeV and $k_\perp$ is a few GeV, typically the region $x\sim \xi \sim 10^{-5}$ of the proton wavefunction is probed. At such small values of $x$, one may have to switch to the Regge limit calculation \cite{Fucilla:2025kit} including the gluon saturation  \cite{Gelis:2010nm}. The two calculations are qualitatively similar, though. In particular, the peak structure around $k_\perp\sim  m$ and $z\sim \frac{1}{2}$ was also  found  in  \cite{Fucilla:2025kit}. It would be interesting to perform a detailed comparison of the spin density matrix with or without gluon saturation  and small-$x$ evolution effects.  

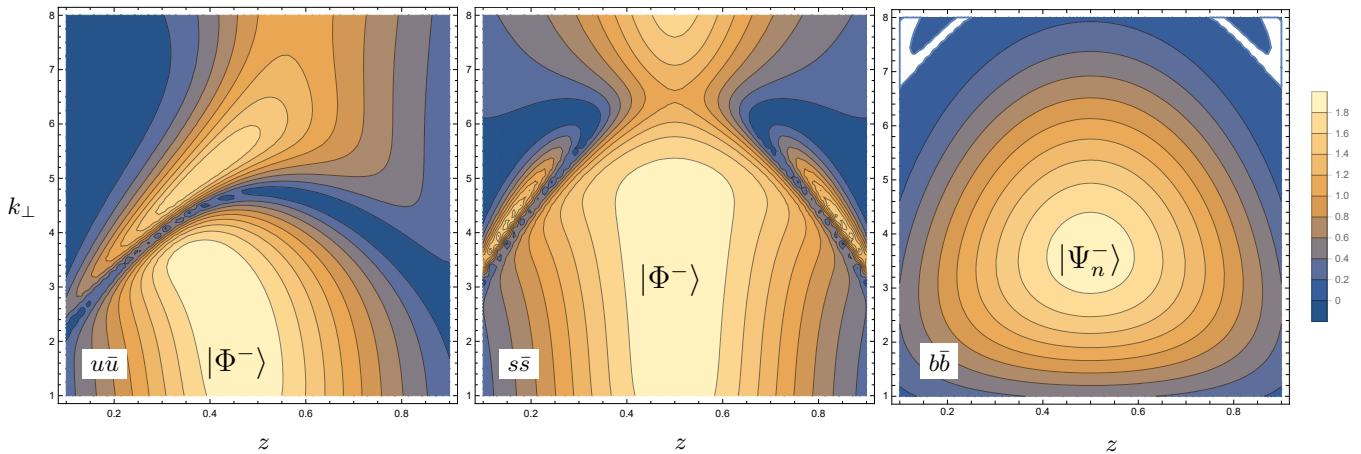
\begin{figure}[t]
    \centering
\hspace{5mm}
    \begin{minipage}{0.3\textwidth}
    \begin{overpic}[width=\textwidth]{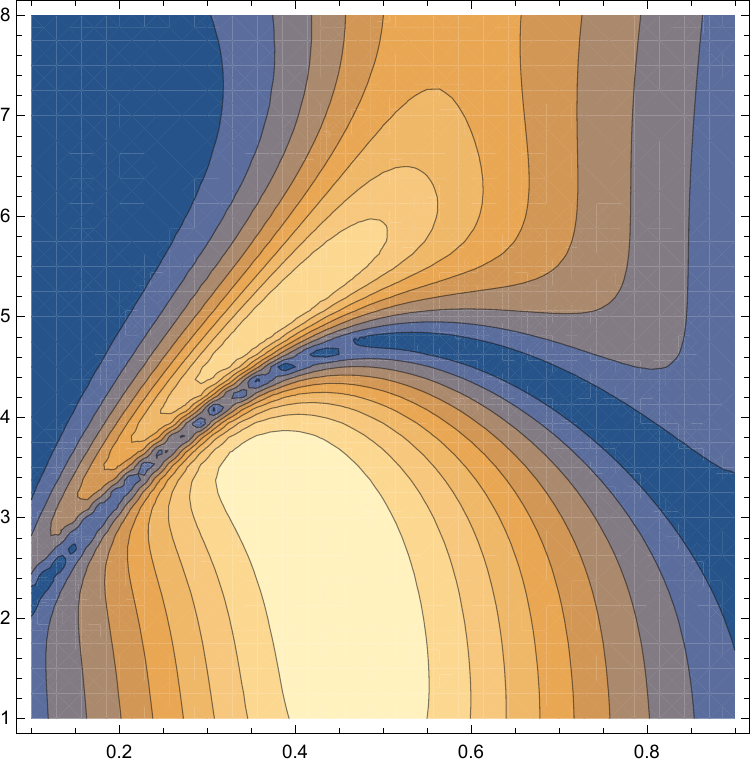}
     \put(8,12){\colorbox{white}{\textbf{$u\bar{u}$}}}
            \put(50,-7){$z$}
            \put(33,12){ \large{ $|\Phi^-\rangle$}}
            \put(-10,50){$k_\perp$}
            \end{overpic}
    \end{minipage}
    \begin{minipage}{0.3\textwidth}
      \begin{overpic}[width=\textwidth]{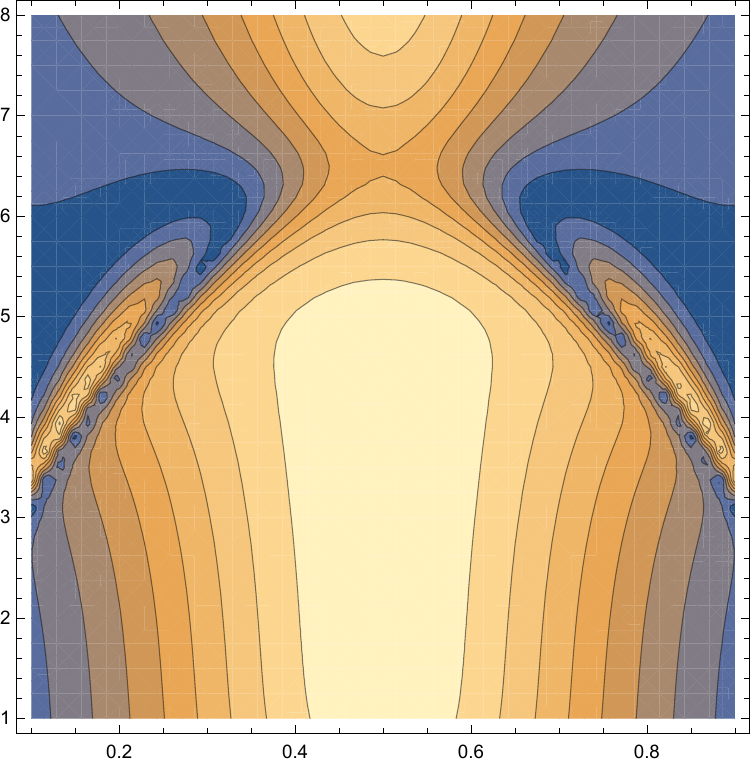}
        \put(8,12){\colorbox{white}{\textbf{$s\bar{s}$}}}
      \put(37,32){ \large{ $|\Phi^-\rangle$}}
            \put(50,-7){$z$}
            \end{overpic}
    \end{minipage}
 \begin{minipage}{0.348\textwidth}
      \begin{overpic}[width=\textwidth]{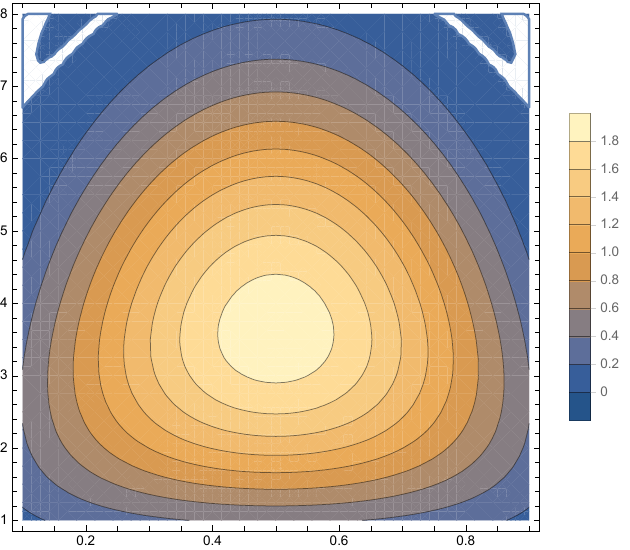}
        \put(8,10){\colorbox{white}{\textbf{$b\bar{b}$}}}
            \put(47,-7){$z$}
              \put(33,32){ \large{ $|\Psi_n^-\rangle$}}
            \end{overpic}
    \end{minipage}
    \vspace{0.02\textwidth}

    \caption[*]{Left hand side (\ref{max})  in electroproduction at $Q=10$ GeV, $W=30$ GeV. Left: $u\bar{u}$ pair, Middle: $s\bar{s}$ pair, Right: $b\bar{b}$ pair. The pair is not entangled in the white region.  }
    \label{2}
\end{figure}

Next, we consider electroproduction at $W=30$ GeV and $Q=10$ GeV. In Fig.~\ref{2}, we plot the left hand side of (\ref{max}) (with $T\to {\rm DIS}$ in the superscript) for $u\bar{u}$ (left), $s\bar{s}$ (middle), $b\bar{b}$ (right) pairs. The $u\bar{u}$ plot is asymmetric under $z\leftrightarrow \bar{z}$. This is because the valence quark integrals $I_L^q$, $I_T^{q_{1,2}}$ with $q=u,d$ do not have this symmetry. Actually, the same is true in  Fig.~\ref{1} (left), but the asymmetry is barely noticeable because at high energy the cross section is dominated by the gluon contribution which is symmetric.  Since $Q$ is relatively high, the results reflect a significant interference between the transverse and longitudinal photon contributions.  In $b\bar{b}$ (and $c\bar{c}$) production, we find  narrow white regions where the pair is not entangled (i.e., the pair is separable). We have studied other sets of $(W,Q)$ and found that regions of no-entanglement appear only for $c\bar{c},b\bar{b}$ pairs  in small patches at most. This means that, even though  the conditions (\ref{id1}) do not hold away from the Regge limit, entanglement remains a robust feature of exclusive $q\bar{q}$ production.

\subsection{Bell nonlocality}

\begin{figure}[t]
    \centering
\hspace{3mm}
    \begin{minipage}{0.3\textwidth}
    \begin{overpic}[width=\textwidth]{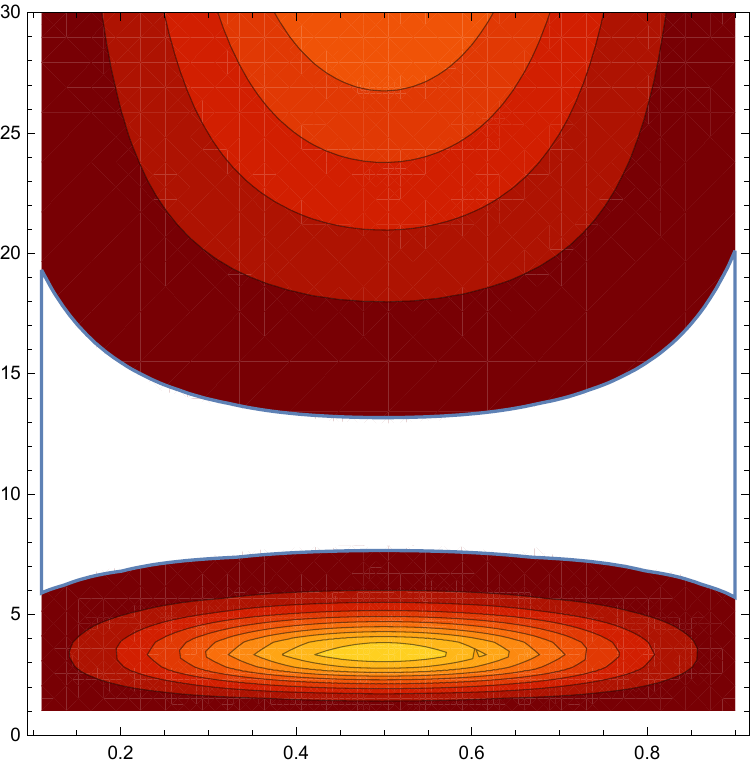}
     \put(8,90){\colorbox{white}{\textbf{$b\bar{b}$}}}
            \put(50,-7){$z$}
            \put(-8,50){$k_\perp$}
            \end{overpic}
    \end{minipage}
    \hspace{2mm}
    \begin{minipage}{0.3\textwidth}
      \begin{overpic}[width=\textwidth]{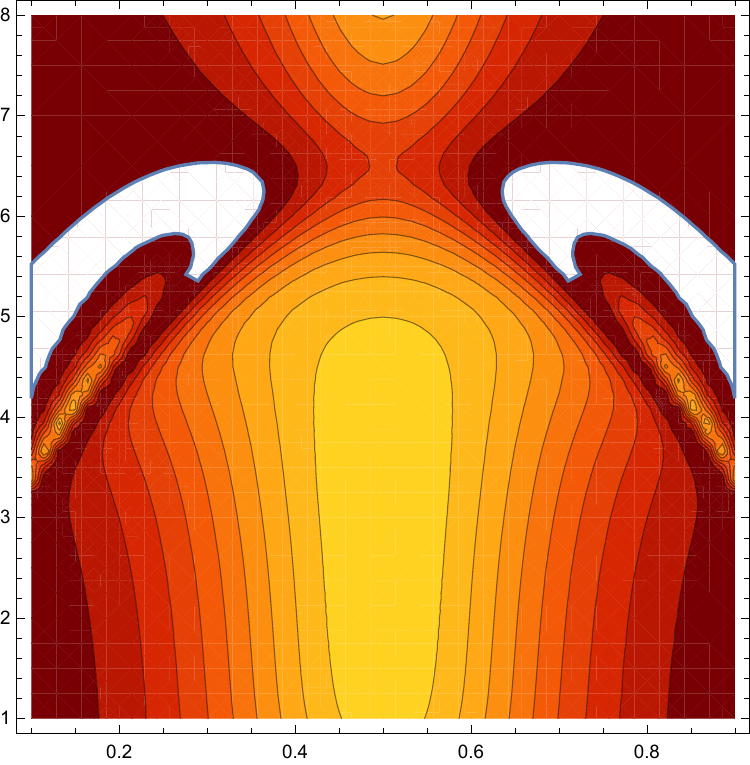}
       \put(8,90){\colorbox{white}{\textbf{$s\bar{s}$}}}
            \put(50,-7){$z$}
            \end{overpic}
    \end{minipage}
 \begin{minipage}{0.348\textwidth}
 \vspace{-1mm}
      \begin{overpic}[width=\textwidth]{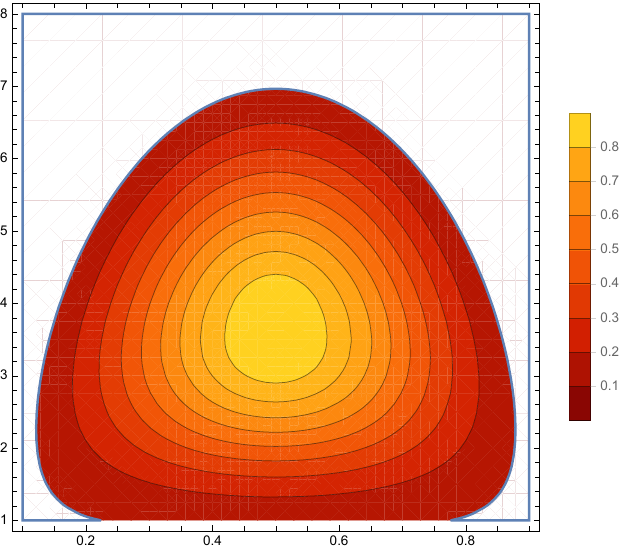}
       \put(8,78){$b\bar{b}$}
           \put(50,-7){$z$}
            \end{overpic}
    \end{minipage}
    \vspace{0.02\textwidth}

    \caption[*]{Plot of $\mu_1+\mu_2-1$, see (\ref{eigen}). The Bell-CHSH inequality is not violated in the white regions. Left: $b\bar{b}$ production in UPC at $W=500$ GeV, $Q=0$, compare with Fig.~\ref{1} (right). Middle: $s\bar{s}$ production in electroproduction, compare with Fig.~\ref{2} (middle).  Right: $b\bar{b}$ pair production in electroproduction, compare with Fig.~\ref{2} (right).   }
    \label{chsh}
\end{figure}
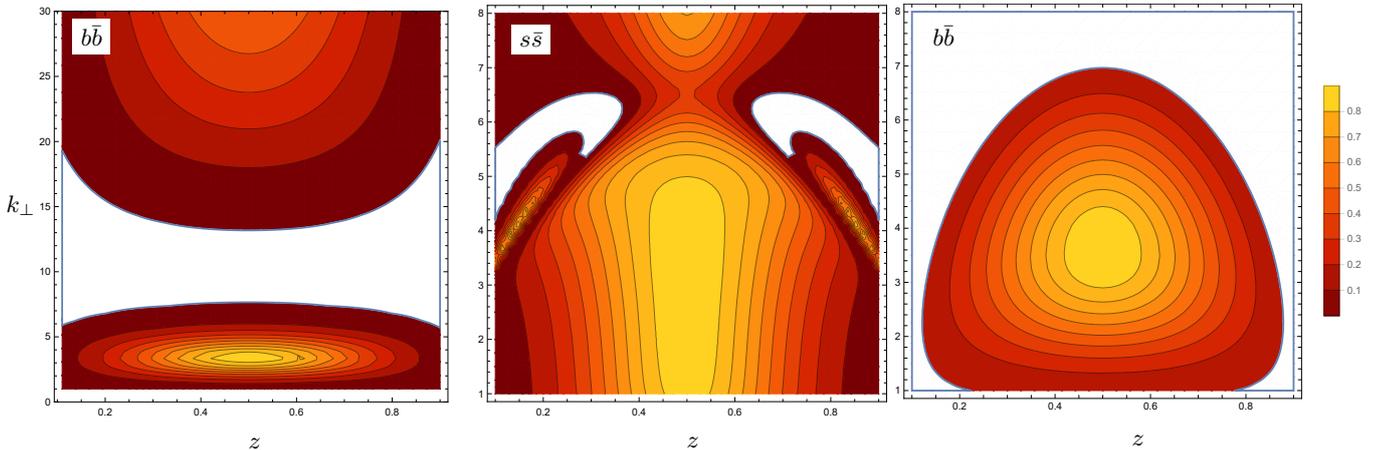

Next, we delineate the regions where the $q\bar{q}$ pair exhibits Bell nonlocality. In Fig.~\ref{chsh} (left),  we plot $\mu_1+\mu_2-1$ (see (\ref{eigen})) for $b\bar{b}$ pairs in the same UPC kinematics as in Fig.~\ref{1} (right). We see a large white region where $\mu_1+\mu_2-1<0$, meaning that the Bell-CHSH inequality is not violated there. This is in contrast to the finding in \cite{Fucilla:2025kit} that, in the Regge limit, the pair exhibits Bell nonlocality in the entire kinematical region of $(z,k_\perp)$. It is also in contrast to Fig.~\ref{1} (right) which shows that the pair remains entangled even away from the Regge limit. These observations are consistent with the fact that, in general, Bell nonlocality is a sufficient but not necessary condition for entanglement. Since the violation of the Bell-CHSH inequality requires stronger quantum correlations than entanglement, it can be seen in a narrower region of phase space. 

In the middle and right panels of Fig.~\ref{chsh}, we plot the same quantity  for the $s\bar{s}$ and $b\bar{b}$ pairs in electroproduction ($W=30$ GeV, $Q=10$ GeV) to be directly compared to the middle and right panels in  Fig.~\ref{2}. Again we find regions where Bell nonlocality is lost, even though the pairs are still entangled.

\subsection{Magic}

\begin{figure}[t]
    \centering

    \begin{minipage}{0.3\textwidth}
    \begin{overpic}[width=\textwidth]{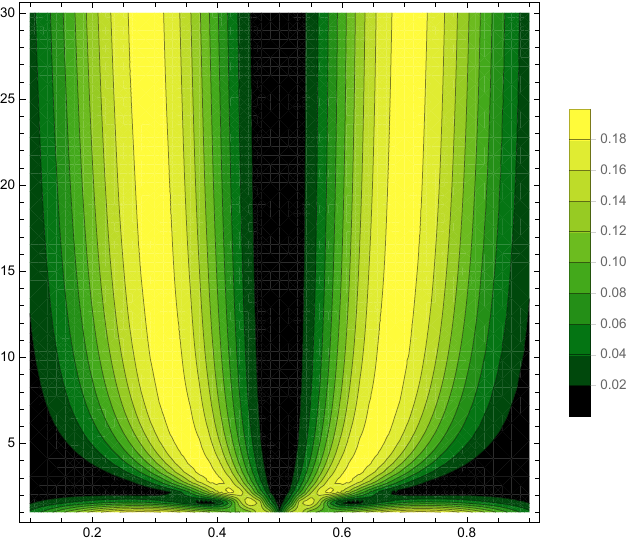}
   \put(8,76){\colorbox{white}{\textbf{$u\bar{u}$}}}
            \put(43,-7){$z$}
            \put(38,60){{\color{white} $|\Phi^-\rangle$}}
            \put(-10,50){$k_\perp$}
            \end{overpic}
    \end{minipage}
    \begin{minipage}{0.3\textwidth}
      \begin{overpic}[width=\textwidth]{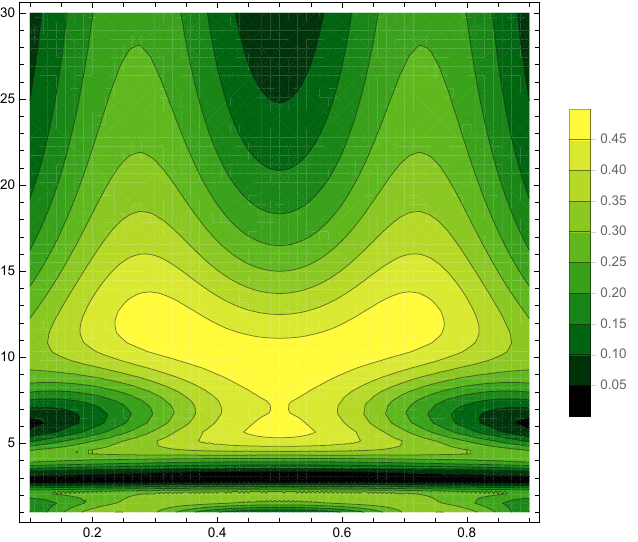}
       \put(8,76){\colorbox{white}{\textbf{$c\bar{c}$}}}
            \put(43,-7){$z$}
            \end{overpic}
    \end{minipage}
 \begin{minipage}{0.3\textwidth}
 \vspace{-1mm}
      \begin{overpic}[width=\textwidth]{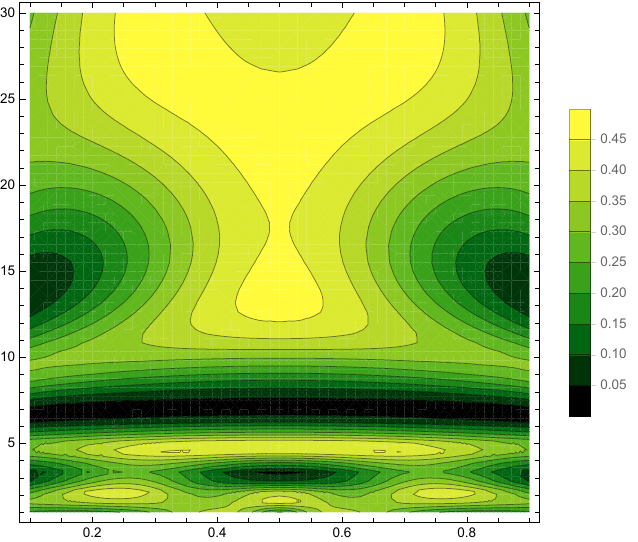}
       \put(8,76){\colorbox{white}{\textbf{$b\bar{b}$}}}
           \put(43,-7){$z$}
              \put(36,9){ {\color{white} $|\Psi_n^-\rangle$}}
            \end{overpic}
    \end{minipage}
    \vspace{0.02\textwidth}

    \caption[*]{Stabilizer R\'enyi entropy (\ref{stabi}) in UPC ($W=500$ GeV) for $u\bar{u}$ (left), $c\bar{c}$ (middle), and $b\bar{b}$ (right) pairs.   }
    \label{magi1}
\end{figure}

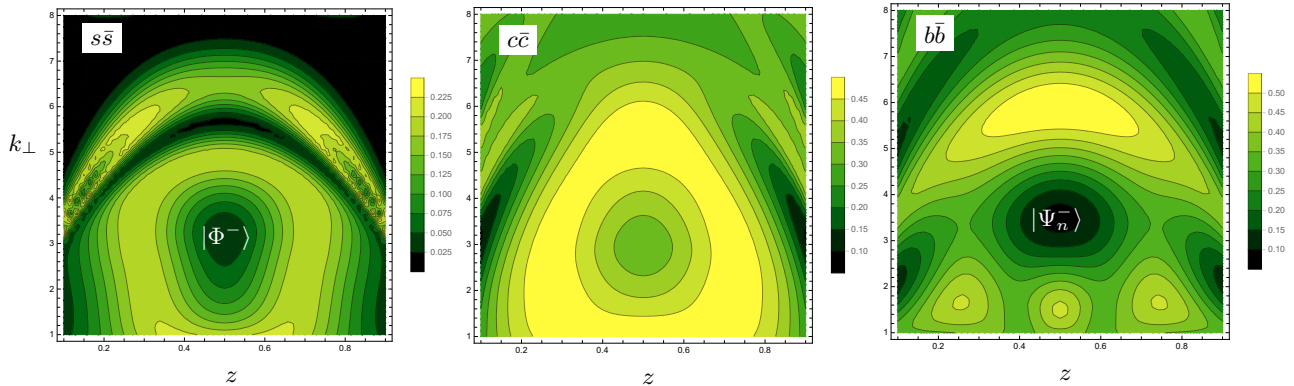
\begin{figure}[t]
    \centering

    \begin{minipage}{0.3\textwidth}
    \begin{overpic}[width=\textwidth]{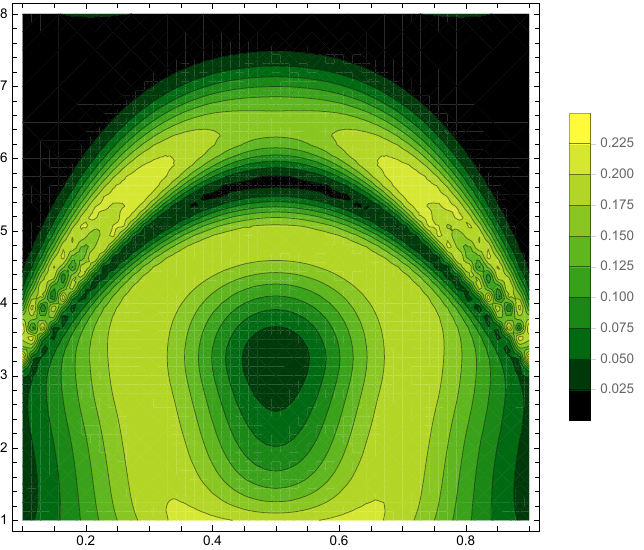}
     \put(8,76){\colorbox{white}{\textbf{$s\bar{s}$}}}
            \put(43,-7){$z$}
            \put(-10,50){$k_\perp$}  \put(37,27){{\color{white} $|\Phi^-\rangle$}}
            \end{overpic}
    \end{minipage}
    \begin{minipage}{0.3\textwidth}
      \begin{overpic}[width=\textwidth]{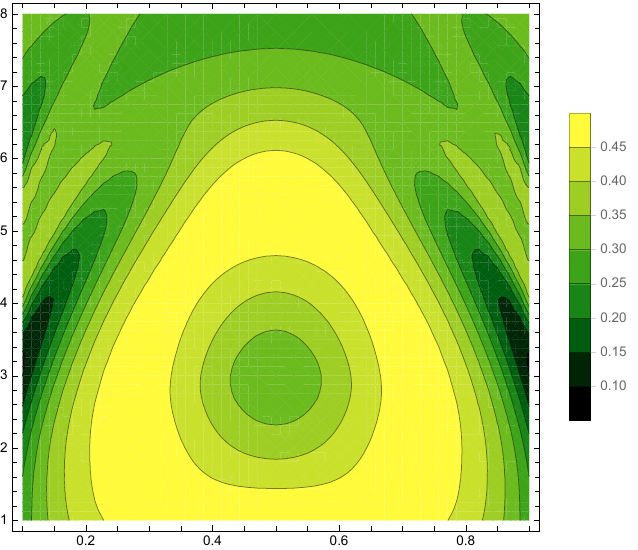}
       \put(8,76){\colorbox{white}{\textbf{$c\bar{c}$}}}
            \put(43,-7){$z$}
            \end{overpic}
    \end{minipage}
 \begin{minipage}{0.3\textwidth}
 \vspace{-1mm}
      \begin{overpic}[width=\textwidth]{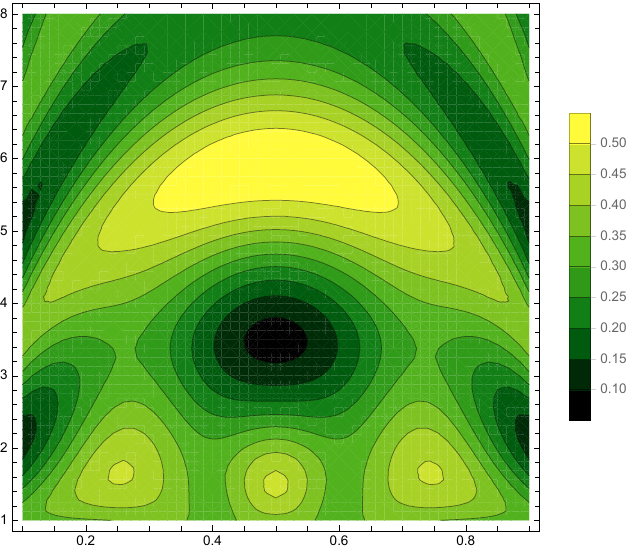}
       \put(8,76){\colorbox{white}{\textbf{$b\bar{b}$}}}
           \put(43,-7){$z$}
              \put(34,31){ {\color{white} $|\Psi_n^-\rangle$}}
            \end{overpic}
    \end{minipage}
    \vspace{0.02\textwidth}

    \caption[*]{Stabilizer R\'enyi entropy (\ref{stabi}) in electroproduction  ($W=30$ GeV, $Q=10$ GeV) for $s\bar{s}$ (left), $c\bar{c}$ (middle), and $b\bar{b}$ (right) pairs.   }
    \label{magi2}
\end{figure}

Finally, in Fig.~\ref{magi1} and Fig.~\ref{magi2}, we plot the stabilizer R\'enyi entropy $M_2$ (\ref{stabi}) in UPC and electroproduction, respectively, for different quark flavors.  As expected, $M_2$ is suppressed when entanglement is nearly maximal. But other than this general statement, it is difficult to anticipate in which kinematical regions $M_2$ is enhanced. We however see that $M_2$ tends to be larger for heavier quarks, with the maximal value  around $0.58$ for the $b\bar{b}$ pair in electroproduction with $W=30$ GeV, $Q=3$ GeV, see Fig.~\ref{magi2} (right). Curiously, in the entire parameter  space we have explored, $M_2$ is always  smaller than (\ref{minimum}), although in principle it can exceed this bound because $B_n,\bar{B}_n\neq 0$.

\section{Conclusions}

In this paper, we have analyzed exclusive quark-antiquark production in electroproduction and UPCs from the viewpoint of Quantum Information Science (QIS). This is a generalization of the previous work in the high energy (Regge) limit \cite{Fucilla:2025kit} to more practical CM energies at the EIC where collinear factorization with GPDs is an appropriate theoretical tool.  The scattering amplitudes computed in the collinear factorization framework are complex, in contrast to being purely real in the one-gluon exchange approximation  \cite{Qi:2025onf} and purely imaginary in the Regge limit.  An immediate consequence  is that massive $q\bar{q}$ pairs are transversely polarized without the usual $\alpha_s$ suppression in inclusive production. In low-energy UPC or electroproduction, the polarization can easily exceed 50\% in certain kinematic regions.  Moreover, we have observed rich patterns of  entanglement, Bell nonlocality and magic due to the interference  between the real and imaginary parts of the scattering amplitudes, and also between the transverse and longitudinal photons. Together with the EIC's ability to vary $W$ and $Q$ over an unprecedentedly  wide kinematical range, this makes DIS a particularly fascinating arena for exploring QIS in collider physics. 


We have focused on the polarization and  entanglement of the produced $q\bar{q}$ pairs  to leading order  in perturbation theory. In future,  the impact of the next-to-leading order corrections should  be investigated (cf.  \cite{Boussarie:2016ogo}). Moreover, other types of GPDs such as the polarized GPDs $\tilde{H}_{q,g}$ can contribute and become important at low energy. 
Most importantly, whether the entanglement  properties discussed in this paper survive  as observables in realistic experiments is a nontrivial problem that requires separate theoretical investigations.  For massive quarks $s,c,b$ (or antiquarks $\bar{s},\bar{c},\bar{b}$), the canonical method is to tag events where the quarks fragment into heavy baryons $\Lambda,\Lambda_c,\Lambda_b$ (or antibaryons $\bar{\Lambda},\bar{\Lambda}_c,\bar{\Lambda}_b$) \cite{Galanti:2015pqa}. Their subsequent weak decays such as $\Lambda \to p+\pi^-$ allows one to  reconstruct the polarization vectors $B_a,\bar{B}_a$ \cite{Bunce:1976yb,Lundberg:1989hw}  and the spin density matrix $C_{ab}$ \cite{Tornqvist:1980af,Afik:2025grr}. For light quarks,  methods outlined in  \cite{Cheng:2025cuv} may be applicable to DIS. In both cases, detailed simulations with a realistic detector setup at the EIC are necessary \cite{Lin:2025eci,Gu:2025ijz}.

\section*{Acknowledgments}

The authors were supported by the U.S. Department
of Energy under Contract No. DE-SC0012704, and also by LDRD funds from Brookhaven Science Associates  and  the framework of the Saturated Glue (SURGE) Topical Theory Collaboration.

\appendix
\section{Maximal magic without polarization}

In this Appendix, we determine the upper limit of the stabilizer R\'enyi entropy \cite{Leone:2021rzd} for generic two qubit systems with vanishing polarization.  
Following \cite{Liu:2025frx}, we parametrize  pure two-qubit states  as 
\beq
|\psi\rangle = c_1|++\rangle +c_2|+-\rangle + c_3|-+\rangle +c_4|--\rangle,
\eeq
\beq
c_1=\sin\theta_1\sin\theta_2e^{i\phi_1}, \qquad c_2=\sin\theta_1\cos\theta_2e^{i\phi_2}, \qquad 
c_3=\cos\theta_1\sin\theta_3e^{i\phi_3}, \qquad c_4=\cos\theta_1\cos\theta_3. \label{c14}
\eeq
Without loss of generality, we may take $0\le \theta_{1,2,3}\le \frac{\pi}{2}$ and $0\le \phi_{1,2,3}\le 2\pi$. 
The polarization of the first qubit can be read off from the reduced density matrix traced over the second qubit   
\beq
{\rm Tr}_2|\psi\rangle\langle \psi| =\begin{pmatrix} |c_1|^2+|c_2|^2 & c_1c_3^*+c_2c_4^* \\ c_1^*c_3+c_2^*c_4 & |c_3|^2+|c_4|^2\end{pmatrix} = \frac{1+\vec{B}_1\cdot  \vec{\sigma}}{2}.
\eeq
If we require that $\vec{B}_1=0$, then  
\beq
|c_1|^2+|c_2|^2=|c_3|^2+|c_4|^2 =\frac{1}{2}, \qquad c_1c_3^*+c_2c_4^*=0. \label{b1}
\eeq
Similarly, if we require that $\vec{B}_2=0$, 
\beq
|c_1|^2+|c_3|^2=|c_2|^2+|c_4|^2=\frac{1}{2}, \qquad c_1c_2^*+c_3c_4^*=0. \label{b2}
\eeq
Substituting (\ref{c14}) into (\ref{b1}), we get 
\beq
\sin\theta_1^2=\cos\theta_1^2 \qquad \sin\theta_2\sin\theta_3e^{i(\phi_1-\phi_3)}+\cos\theta_2\cos\theta_3e^{i\phi_2}=0.
\eeq
This means that 
\beq
\theta_1=\frac{\pi}{4} \qquad \phi_1-\phi_3=\phi_2\pm \pi, \qquad \theta_3=\frac{\pi}{2}-\theta_2. \label{cond}
\eeq
It is easy to check that when (\ref{cond}) is satisfied, (\ref{b2}) is automatically satisfied. 
The maximal value of $M_2$ for systems with  $\vec{B}_1=\vec{B}_2=0$ is therefore obtained by minimizing  (16) of \cite{Liu:2025frx} with the constraints  (\ref{cond}). There are many degenerate minima. For example, we find  
\beq
\cos\theta_2=\sqrt{\frac{1}{3}},
\qquad 
\phi_1=
\frac{\pi}{2}, \qquad \phi_2=
\frac{5}{4}\pi, \qquad \phi_3=\frac{\pi}{4},
\eeq
to a good numerical accuracy. 
The corresponding value of $M_2$ is given by (\ref{minimum}). 

Now let us further require that $\theta_2=\frac{\pi}{2}$, and accordingly, $\theta_3=0$. (Alternatively, one may set $\theta_2=0$ and $\theta_3=\frac{\pi}{2}$.) The density matrix becomes 
\beq
\rho^L= \frac{1}{2}\begin{pmatrix} 1 & 0 & 0 & e^{i\phi_1} \\ 0 & 0 & 0 & 0 \\ 0 & 0 & 0 & 0 \\ 
 e^{-i\phi_1} & 0 & 0 & 1\end{pmatrix}.
 \eeq
Minimizing (16) of \cite{Liu:2025frx} under this additional constraint, we find the  extremal condition $\cos(4\phi_1)=-1$, or equivalently,  
\beq
\phi_1=\frac{\pi}{4},\frac{3\pi}{4},\frac{5\pi}{4},\frac{7\pi}{4}. \label{four}
\eeq
At these points, $M_2$ takes the maximal value which coincides with (\ref{l2max}). 
The four values of $\phi_1$ in (\ref{four})  correspond to the four branches  (\ref{maxrho}).

\bibliography{ref}

\begin{thebibliography}{75}
\providecommand{\natexlab}[1]{#1}
\providecommand{\url}[1]{\texttt{#1}}
\expandafter\ifx\csname urlstyle\endcsname\relax
  \providecommand{\doi}[1]{doi: #1}\else
  \providecommand{\doi}{doi: \begingroup \urlstyle{rm}\Url}\fi

\bibitem[Wu and Shaknov(1950)]{Wu:1950zz}
C.~S. Wu and I.~Shaknov.
\newblock {The Angular Correlation of Scattered Annihilation Radiation}.
\newblock \emph{Phys. Rev.}, 77:\penalty0 136--136, 1950.
\newblock \doi{10.1103/PhysRev.77.136}.

\bibitem[Bell(1964)]{Bell:1964kc}
J.~S. Bell.
\newblock {On the Einstein-Podolsky-Rosen paradox}.
\newblock \emph{Physics Physique Fizika}, 1:\penalty0 195--200, 1964.
\newblock \doi{10.1103/PhysicsPhysiqueFizika.1.195}.

\bibitem[Clauser et~al.(1969)Clauser, Horne, Shimony, and Holt]{Clauser:1969ny}
John~F. Clauser, Michael~A. Horne, Abner Shimony, and Richard~A. Holt.
\newblock {Proposed experiment to test local hidden variable theories}.
\newblock \emph{Phys. Rev. Lett.}, 23:\penalty0 880--884, 1969.
\newblock \doi{10.1103/PhysRevLett.23.880}.

\bibitem[Aspect et~al.(1981)Aspect, Grangier, and Roger]{Aspect:1981zz}
Alain Aspect, Philippe Grangier, and Gerard Roger.
\newblock {Experimental Tests of Realistic Local Theories via Bell's Theorem}.
\newblock \emph{Phys. Rev. Lett.}, 47:\penalty0 460--6443, 1981.
\newblock \doi{10.1103/PhysRevLett.47.460}.

\bibitem[Weihs et~al.(1998)Weihs, Jennewein, Simon, Weinfurter, and Zeilinger]{Weihs:1998gy}
Gregor Weihs, Thomas Jennewein, Christoph Simon, Harald Weinfurter, and Anton Zeilinger.
\newblock {Violation of Bell's inequality under strict Einstein locality conditions}.
\newblock \emph{Phys. Rev. Lett.}, 81:\penalty0 5039--5043, 1998.
\newblock \doi{10.1103/PhysRevLett.81.5039}.

\bibitem[Bernreuther and Brandenburg(1994)]{Bernreuther:1993hq}
Werner Bernreuther and Arnd Brandenburg.
\newblock {Tracing CP violation in the production of top quark pairs by multiple TeV proton proton collisions}.
\newblock \emph{Phys. Rev. D}, 49:\penalty0 4481--4492, 1994.
\newblock \doi{10.1103/PhysRevD.49.4481}.

\bibitem[Baumgart and Tweedie(2013)]{Baumgart:2012ay}
Matthew Baumgart and Brock Tweedie.
\newblock {A New Twist on Top Quark Spin Correlations}.
\newblock \emph{JHEP}, 03:\penalty0 117, 2013.
\newblock \doi{10.1007/JHEP03(2013)117}.

\bibitem[Bernreuther et~al.(2015)Bernreuther, Heisler, and Si]{Bernreuther:2015yna}
Werner Bernreuther, Dennis Heisler, and Zong-Guo Si.
\newblock {A set of top quark spin correlation and polarization observables for the LHC: Standard Model predictions and new physics contributions}.
\newblock \emph{JHEP}, 12:\penalty0 026, 2015.
\newblock \doi{10.1007/JHEP12(2015)026}.

\bibitem[Afik and de~Nova(2021)]{Afik:2020onf}
Yoav Afik and Juan Ram{\'o}n~Mu{\~n}oz de~Nova.
\newblock {Entanglement and quantum tomography with top quarks at the LHC}.
\newblock \emph{Eur. Phys. J. Plus}, 136\penalty0 (9):\penalty0 907, 2021.
\newblock \doi{10.1140/epjp/s13360-021-01902-1}.

\bibitem[Dong et~al.(2024)Dong, Gon{\c{c}}alves, Kong, and Navarro]{Dong:2023xiw}
Zhongtian Dong, Dorival Gon{\c{c}}alves, Kyoungchul Kong, and Alberto Navarro.
\newblock {Entanglement and Bell inequalities with boosted tt{\textasciimacron}}.
\newblock \emph{Phys. Rev. D}, 109\penalty0 (11):\penalty0 115023, 2024.
\newblock \doi{10.1103/PhysRevD.109.115023}.

\bibitem[Aad et~al.(2024)]{ATLAS:2023fsd}
Georges Aad et~al.
\newblock {Observation of quantum entanglement with top quarks at the ATLAS detector}.
\newblock \emph{Nature}, 633\penalty0 (8030):\penalty0 542--547, 2024.
\newblock \doi{10.1038/s41586-024-07824-z}.

\bibitem[Hayrapetyan et~al.(2024{\natexlab{a}})]{CMS:2024pts}
Aram Hayrapetyan et~al.
\newblock {Observation of quantum entanglement in top quark pair production in proton{\textendash}proton collisions at $\sqrt{s} = 13$ TeV}.
\newblock \emph{Rept. Prog. Phys.}, 87\penalty0 (11):\penalty0 117801, 2024{\natexlab{a}}.
\newblock \doi{10.1088/1361-6633/ad7e4d}.

\bibitem[Hayrapetyan et~al.(2024{\natexlab{b}})]{CMS:2024zkc}
Aram Hayrapetyan et~al.
\newblock {Measurements of polarization and spin correlation and observation of entanglement in top quark pairs using lepton+jets events from proton-proton collisions at s=13{\,}{\,}TeV}.
\newblock \emph{Phys. Rev. D}, 110\penalty0 (11):\penalty0 112016, 2024{\natexlab{b}}.
\newblock \doi{10.1103/PhysRevD.110.112016}.

\bibitem[Sirunyan et~al.(2019)]{CMS:2019nrx}
Albert~M Sirunyan et~al.
\newblock {Measurement of the top quark polarization and $\mathrm{t\bar{t}}$ spin correlations using dilepton final states in proton-proton collisions at $\sqrt{s} =$ 13 TeV}.
\newblock \emph{Phys. Rev. D}, 100\penalty0 (7):\penalty0 072002, 2019.
\newblock \doi{10.1103/PhysRevD.100.072002}.

\bibitem[Barr et~al.(2024)Barr, Fabbrichesi, Floreanini, Gabrielli, and Marzola]{Barr:2024djo}
Alan~J. Barr, Marco Fabbrichesi, Roberto Floreanini, Emidio Gabrielli, and Luca Marzola.
\newblock {Quantum entanglement and Bell inequality violation at colliders}.
\newblock \emph{Prog. Part. Nucl. Phys.}, 139:\penalty0 104134, 2024.
\newblock \doi{10.1016/j.ppnp.2024.104134}.

\bibitem[Afik et~al.(2025{\natexlab{a}})]{Afik:2025ejh}
Yoav Afik et~al.
\newblock {Quantum Information meets High-Energy Physics: Input to the update of the European Strategy for Particle Physics}.
\newblock 3 2025{\natexlab{a}}.

\bibitem[Aguilar-Saavedra(2023)]{Aguilar-Saavedra:2022mpg}
J.~A. Aguilar-Saavedra.
\newblock {Laboratory-frame tests of quantum entanglement in H{\textrightarrow}WW}.
\newblock \emph{Phys. Rev. D}, 107\penalty0 (7):\penalty0 076016, 2023.
\newblock \doi{10.1103/PhysRevD.107.076016}.

\bibitem[Fabbrichesi et~al.(2023)Fabbrichesi, Floreanini, Gabrielli, and Marzola]{Fabbrichesi:2023cev}
Marco Fabbrichesi, Roberto Floreanini, Emidio Gabrielli, and Luca Marzola.
\newblock {Bell inequalities and quantum entanglement in weak gauge boson production at the LHC and future colliders}.
\newblock \emph{Eur. Phys. J. C}, 83\penalty0 (9):\penalty0 823, 2023.
\newblock \doi{10.1140/epjc/s10052-023-11935-8}.

\bibitem[Altakach et~al.(2023)Altakach, Lamba, Maltoni, Mawatari, and Sakurai]{Altakach:2022ywa}
Mohammad~Mahdi Altakach, Priyanka Lamba, Fabio Maltoni, Kentarou Mawatari, and Kazuki Sakurai.
\newblock {Quantum information and CP measurement in H{\textrightarrow}{\ensuremath{\tau}}+{\ensuremath{\tau}}- at future lepton colliders}.
\newblock \emph{Phys. Rev. D}, 107\penalty0 (9):\penalty0 093002, 2023.
\newblock \doi{10.1103/PhysRevD.107.093002}.

\bibitem[Han et~al.(2025)Han, Low, and Su]{Han:2025ewp}
Tao Han, Matthew Low, and Youle Su.
\newblock {Entanglement and Bell Nonlocality in $\tau^+ \tau^-$ at the BEPC}.
\newblock 1 2025.

\bibitem[Kats and Uzan(2024)]{Kats:2023zxb}
Yevgeny Kats and David Uzan.
\newblock {Prospects for measuring quark polarization and spin correlations in $b\overline{b }$ and $c\overline{c }$ samples at the LHC}.
\newblock \emph{JHEP}, 03:\penalty0 063, 2024.
\newblock \doi{10.1007/JHEP03(2024)063}.

\bibitem[Afik et~al.(2025{\natexlab{b}})Afik, Kats, de~Nova, Soffer, and Uzan]{Afik:2025grr}
Yoav Afik, Yevgeny Kats, Juan Ram{\'o}n~Mu{\~n}oz de~Nova, Abner Soffer, and David Uzan.
\newblock {Entanglement and Bell nonlocality with bottom-quark pairs at hadron colliders}.
\newblock \emph{Phys. Rev. D}, 111\penalty0 (11):\penalty0 L111902, 2025{\natexlab{b}}.
\newblock \doi{10.1103/fhkc-kfhr}.

\bibitem[Cheng and Yan(2025)]{Cheng:2025cuv}
Kun Cheng and Bin Yan.
\newblock {Bell Inequality Violation of Light Quarks in Dihadron Pair Production at Lepton Colliders}.
\newblock \emph{Phys. Rev. Lett.}, 135\penalty0 (1):\penalty0 011902, 2025.
\newblock \doi{10.1103/gmqz-v4cl}.

\bibitem[Fabbrichesi et~al.(2021)Fabbrichesi, Floreanini, and Panizzo]{Fabbrichesi:2021npl}
M.~Fabbrichesi, R.~Floreanini, and G.~Panizzo.
\newblock {Testing Bell Inequalities at the LHC with Top-Quark Pairs}.
\newblock \emph{Phys. Rev. Lett.}, 127\penalty0 (16):\penalty0 161801, 2021.
\newblock \doi{10.1103/PhysRevLett.127.161801}.

\bibitem[Severi et~al.(2022)Severi, Boschi, Maltoni, and Sioli]{Severi:2021cnj}
Claudio Severi, Cristian Degli~Esposti Boschi, Fabio Maltoni, and Maximiliano Sioli.
\newblock {Quantum tops at the LHC: from entanglement to Bell inequalities}.
\newblock \emph{Eur. Phys. J. C}, 82\penalty0 (4):\penalty0 285, 2022.
\newblock \doi{10.1140/epjc/s10052-022-10245-9}.

\bibitem[Abel et~al.(1992)Abel, Dittmar, and Dreiner]{Abel:1992kz}
S.~A. Abel, M.~Dittmar, and Herbert~K. Dreiner.
\newblock {Testing locality at colliders via Bell's inequality?}
\newblock \emph{Phys. Lett. B}, 280:\penalty0 304--312, 1992.
\newblock \doi{10.1016/0370-2693(92)90071-B}.

\bibitem[Abel et~al.(2025)Abel, Dreiner, Sengupta, and Ubaldi]{Abel:2025skj}
Steven~A. Abel, Herbi~K. Dreiner, Rhitaja Sengupta, and Lorenzo Ubaldi.
\newblock {Colliders are Testing neither Locality via Bell's Inequality nor Entanglement versus Non-Entanglement}.
\newblock 7 2025.

\bibitem[Low(2025)]{Low:2025aqq}
Matthew Low.
\newblock {Addressing local realism through Bell tests at colliders}.
\newblock \emph{Phys. Rev. D}, 112\penalty0 (9):\penalty0 096008, 2025.
\newblock \doi{10.1103/15c3-mg5l}.

\bibitem[Qi et~al.(2025)Qi, Guo, and Xiao]{Qi:2025onf}
Wei Qi, Zijing Guo, and Bo-Wen Xiao.
\newblock {Studying Maximal Entanglement and Bell Nonlocality at an Electron-Ion Collider}.
\newblock 6 2025.

\bibitem[Fucilla and Hatta(2026)]{Fucilla:2025kit}
Michael Fucilla and Yoshitaka Hatta.
\newblock {Spin-spin entanglement in diffractive heavy-quark production}.
\newblock \emph{Phys. Rev. D}, 113\penalty0 (3):\penalty0 L031504, 2026.
\newblock \doi{10.1103/gbk8-z3dd}.

\bibitem[Abdul~Khalek et~al.(2022)]{AbdulKhalek:2021gbh}
R.~Abdul~Khalek et~al.
\newblock {Science Requirements and Detector Concepts for the Electron-Ion Collider}: {EIC Yellow Report}.
\newblock \emph{Nucl. Phys. A}, 1026:\penalty0 122447, 2022.
\newblock \doi{10.1016/j.nuclphysa.2022.122447}.

\bibitem[Diehl(2003)]{Diehl:2003ny}
M.~Diehl.
\newblock {Generalized parton distributions}.
\newblock \emph{Phys. Rept.}, 388:\penalty0 41--277, 2003.
\newblock \doi{10.1016/j.physrep.2003.08.002}.

\bibitem[Belitsky and Radyushkin(2005)]{Belitsky:2005qn}
A.~V. Belitsky and A.~V. Radyushkin.
\newblock {Unraveling hadron structure with generalized parton distributions}.
\newblock \emph{Phys. Rept.}, 418:\penalty0 1--387, 2005.
\newblock \doi{10.1016/j.physrep.2005.06.002}.

\bibitem[Leone et~al.(2022)Leone, Oliviero, and Hamma]{Leone:2021rzd}
Lorenzo Leone, Salvatore F.~E. Oliviero, and Alioscia Hamma.
\newblock {Stabilizer R{\'e}nyi Entropy}.
\newblock \emph{Phys. Rev. Lett.}, 128\penalty0 (5):\penalty0 050402, 2022.
\newblock \doi{10.1103/PhysRevLett.128.050402}.

\bibitem[White and White(2024)]{White:2024nuc}
Chris~D. White and Martin~J. White.
\newblock {Magic states of top quarks}.
\newblock \emph{Phys. Rev. D}, 110\penalty0 (11):\penalty0 116016, 2024.
\newblock \doi{10.1103/PhysRevD.110.116016}.

\bibitem[Bunce et~al.(1976)]{Bunce:1976yb}
G.~Bunce et~al.
\newblock {Lambda0 Hyperon Polarization in Inclusive Production by 300-GeV Protons on Beryllium.}
\newblock \emph{Phys. Rev. Lett.}, 36:\penalty0 1113--1116, 1976.
\newblock \doi{10.1103/PhysRevLett.36.1113}.

\bibitem[Lundberg et~al.(1989)]{Lundberg:1989hw}
B.~Lundberg et~al.
\newblock {Polarization in Inclusive $\Lambda$ and $\bar{\Lambda}$ Production at Large $p_T$}.
\newblock \emph{Phys. Rev. D}, 40:\penalty0 3557--3567, 1989.
\newblock \doi{10.1103/PhysRevD.40.3557}.

\bibitem[Airapetian et~al.(2007)]{HERMES:2007fpi}
A.~Airapetian et~al.
\newblock {Transverse Polarization of Lambda and anti-Lambda Hyperons in Quasireal Photoproduction}.
\newblock \emph{Phys. Rev. D}, 76:\penalty0 092008, 2007.
\newblock \doi{10.1103/PhysRevD.76.092008}.

\bibitem[Alexeev et~al.(2022)]{COMPASS:2021bws}
M.~G. Alexeev et~al.
\newblock {Probing transversity by measuring {\ensuremath{\Lambda}} polarisation in SIDIS}.
\newblock \emph{Phys. Lett. B}, 824:\penalty0 136834, 2022.
\newblock \doi{10.1016/j.physletb.2021.136834}.

\bibitem[Braun and Ivanov(2005)]{Braun:2005rg}
V.~M. Braun and D.~Yu. Ivanov.
\newblock {Exclusive diffractive electroproduction of dijets in collinear factorization}.
\newblock \emph{Phys. Rev. D}, 72:\penalty0 034016, 2005.
\newblock \doi{10.1103/PhysRevD.72.034016}.

\bibitem[Jackson(1998)]{Jackson:1998nia}
John~David Jackson.
\newblock \emph{{Classical Electrodynamics}}.
\newblock Wiley, 1998.
\newblock ISBN 978-0-471-30932-1.

\bibitem[Peres and Terno(2004)]{Peres:2002wx}
Asher Peres and Daniel~R. Terno.
\newblock {Quantum information and relativity theory}.
\newblock \emph{Rev. Mod. Phys.}, 76:\penalty0 93--123, 2004.
\newblock \doi{10.1103/RevModPhys.76.93}.

\bibitem[Jacob and Wick(1959)]{Jacob:1959at}
M.~Jacob and G.~C. Wick.
\newblock {On the General Theory of Collisions for Particles with Spin}.
\newblock \emph{Annals Phys.}, 7:\penalty0 404--428, 1959.
\newblock \doi{10.1006/aphy.2000.6022}.

\bibitem[Peskin and Schroeder(1995)]{Peskin:1995ev}
Michael~E. Peskin and Daniel~V. Schroeder.
\newblock \emph{{An Introduction to quantum field theory}}.
\newblock Addison-Wesley, Reading, USA, 1995.
\newblock ISBN 978-0-201-50397-5, 978-0-429-50355-9, 978-0-429-49417-8.
\newblock \doi{10.1201/9780429503559}.

\bibitem[Cheng et~al.(2025)Cheng, Han, and Trifinopoulos]{Cheng:2025zaw}
Kun Cheng, Tao Han, and Sokratis Trifinopoulos.
\newblock {Quantum Information at the Electron-Ion Collider}.
\newblock 10 2025.

\bibitem[Liang and Boros(1997)]{Liang:1997rt}
Zuo-tang Liang and C.~Boros.
\newblock {Hyperon polarization and single spin left-right asymmetry in inclusive production processes at high-energies}.
\newblock \emph{Phys. Rev. Lett.}, 79:\penalty0 3608--3611, 1997.
\newblock \doi{10.1103/PhysRevLett.79.3608}.

\bibitem[Kanazawa and Koike(2001)]{Kanazawa:2000cx}
Y.~Kanazawa and Yuji Koike.
\newblock {Polarization in hadronic Lambda hyperon production and chiral odd twist - three quark distribution}.
\newblock \emph{Phys. Rev. D}, 64:\penalty0 034019, 2001.
\newblock \doi{10.1103/PhysRevD.64.034019}.

\bibitem[Anselmino et~al.(2002)Anselmino, Boer, D'Alesio, and Murgia]{Anselmino:2001js}
M.~Anselmino, Daniel Boer, U.~D'Alesio, and F.~Murgia.
\newblock {Transverse lambda polarization in semiinclusive DIS}.
\newblock \emph{Phys. Rev. D}, 65:\penalty0 114014, 2002.
\newblock \doi{10.1103/PhysRevD.65.114014}.

\bibitem[Zhou et~al.(2008)Zhou, Yuan, and Liang]{Zhou:2008fb}
Jian Zhou, Feng Yuan, and Zuo-Tang Liang.
\newblock {Hyperon Polarization in Unpolarized Scattering Processes}.
\newblock \emph{Phys. Rev. D}, 78:\penalty0 114008, 2008.
\newblock \doi{10.1103/PhysRevD.78.114008}.

\bibitem[Koike et~al.(2017)Koike, Metz, Pitonyak, Yabe, and Yoshida]{Koike:2017fxr}
Yuji Koike, Andreas Metz, Daniel Pitonyak, Kenta Yabe, and Shinsuke Yoshida.
\newblock {Twist-3 fragmentation contribution to polarized hyperon production in unpolarized hadronic collisions}.
\newblock \emph{Phys. Rev. D}, 95\penalty0 (11):\penalty0 114013, 2017.
\newblock \doi{10.1103/PhysRevD.95.114013}.

\bibitem[Kang et~al.(2022)Kang, Terry, Vossen, Xu, and Zhang]{Kang:2021kpt}
Zhong-Bo Kang, John Terry, Anselm Vossen, Qinghua Xu, and Jinlong Zhang.
\newblock {Transverse Lambda production at the future Electron-Ion Collider}.
\newblock \emph{Phys. Rev. D}, 105\penalty0 (9):\penalty0 094033, 2022.
\newblock \doi{10.1103/PhysRevD.105.094033}.

\bibitem[Dharmaratna and Goldstein(1996)]{Dharmaratna:1996xd}
Welathantri G.~D. Dharmaratna and Gary~R. Goldstein.
\newblock {Single quark polarization in quantum chromodynamics subprocesses}.
\newblock \emph{Phys. Rev. D}, 53:\penalty0 1073--1086, 1996.
\newblock \doi{10.1103/PhysRevD.53.1073}.

\bibitem[Beni{\'c} et~al.(2024)Beni{\'c}, Hatta, Kaushik, and Li]{Benic:2024fvk}
Sanjin Beni{\'c}, Yoshitaka Hatta, Abhiram Kaushik, and Hsiang-nan Li.
\newblock {Perturbative QCD contribution to transverse single spin asymmetries in the Drell-Yan process and SIDIS}.
\newblock \emph{Phys. Rev. D}, 109\penalty0 (7):\penalty0 074038, 2024.
\newblock \doi{10.1103/PhysRevD.109.074038}.

\bibitem[Hatta et~al.(2017)Hatta, Xiao, and Yuan]{Hatta:2017cte}
Yoshitaka Hatta, Bo-Wen Xiao, and Feng Yuan.
\newblock {Gluon Tomography from Deeply Virtual Compton Scattering at Small-x}.
\newblock \emph{Phys. Rev. D}, 95\penalty0 (11):\penalty0 114026, 2017.
\newblock \doi{10.1103/PhysRevD.95.114026}.

\bibitem[Nikolaev and Zakharov(1994)]{Nikolaev:1994cd}
Nikolai~N. Nikolaev and B.~G. Zakharov.
\newblock {Splitting the pomeron into two jets: A Novel process at HERA}.
\newblock \emph{Phys. Lett. B}, 332:\penalty0 177--183, 1994.
\newblock \doi{10.1016/0370-2693(94)90876-1}.

\bibitem[Bartels et~al.(1996)Bartels, Lotter, and W{\"u}sthoff]{Bartels:1996ne}
Jochen Bartels, H.~Lotter, and M.~W{\"u}sthoff.
\newblock {Quark-antiquark production in DIS diffractive dissociation}.
\newblock \emph{Phys. Lett. B}, 379:\penalty0 239--248, 1996.
\newblock \doi{10.1016/0370-2693(96)00412-1}.
\newblock [Erratum: Phys.Lett.B 382, 449--449 (1996)].

\bibitem[Peres(1996)]{Peres:1996dw}
Asher Peres.
\newblock {Separability criterion for density matrices}.
\newblock \emph{Phys. Rev. Lett.}, 77:\penalty0 1413--1415, 1996.
\newblock \doi{10.1103/PhysRevLett.77.1413}.

\bibitem[Horodecki(1997)]{Horodecki:1997vt}
Pawel Horodecki.
\newblock {Separability criterion and inseparable mixed states with positive partial transposition}.
\newblock \emph{Phys. Lett. A}, 232:\penalty0 333, 1997.
\newblock \doi{10.1016/S0375-9601(97)00416-7}.

\bibitem[Lang and Caves(2010)]{Lang:2010xtw}
Matthias~D. Lang and Carlton~M. Caves.
\newblock {Quantum Discord and the Geometry of Bell-Diagonal States}.
\newblock \emph{Phys. Rev. Lett.}, 105\penalty0 (15):\penalty0 150501, 2010.
\newblock \doi{10.1103/PhysRevLett.105.150501}.

\bibitem[Horodecki et~al.(1995)Horodecki, Horodecki, and Horodecki]{Horodecki:1995nsk}
R.~Horodecki, P.~Horodecki, and M.~Horodecki.
\newblock {Violating Bell inequality by mixed spin- 1 2 states: necessary and sufficient condition }.
\newblock \emph{Phys. Lett. A}, 200\penalty0 (5):\penalty0 340--344, 1995.
\newblock \doi{10.1016/0375-9601(95)00214-N}.

\bibitem[Werner(1989)]{Werner:1989zz}
Reinhard~F. Werner.
\newblock {Quantum states with Einstein-Podolsky-Rosen correlations admitting a hidden-variable model}.
\newblock \emph{Phys. Rev. A}, 40:\penalty0 4277--4281, 1989.
\newblock \doi{10.1103/PhysRevA.40.4277}.

\bibitem[Gottesman(1998)]{Gottesman:1998hu}
Daniel Gottesman.
\newblock {The Heisenberg representation of quantum computers}.
\newblock In \emph{{22nd International Colloquium on Group Theoretical Methods in Physics}}, pages 32--43, 7 1998.

\bibitem[Liu et~al.(2025{\natexlab{a}})Liu, Low, and Yin]{Liu:2025qfl}
Qiaofeng Liu, Ian Low, and Zhewei Yin.
\newblock {Quantum Magic in Quantum Electrodynamics}.
\newblock 3 2025{\natexlab{a}}.

\bibitem[Gargalionis et~al.(2025)Gargalionis, Moynihan, Trifinopoulos, Wallace, White, and White]{Gargalionis:2025iqs}
John Gargalionis, Nathan Moynihan, Sokratis Trifinopoulos, Ewan N.~V. Wallace, Chris~D. White, and Martin~J. White.
\newblock {Spin versus Magic: Lessons from Gluon and Graviton Scattering}.
\newblock 8 2025.

\bibitem[Liu et~al.(2025{\natexlab{b}})Liu, Low, and Yin]{Liu:2025frx}
Qiaofeng Liu, Ian Low, and Zhewei Yin.
\newblock {Maximal Magic for Two-qubit States}.
\newblock 2 2025{\natexlab{b}}.

\bibitem[Goloskokov and Kroll(2007)]{Goloskokov:2006hr}
S.~V. Goloskokov and P.~Kroll.
\newblock {The Longitudinal cross-section of vector meson electroproduction}.
\newblock \emph{Eur. Phys. J. C}, 50:\penalty0 829--842, 2007.
\newblock \doi{10.1140/epjc/s10052-007-0228-4}.

\bibitem[Goloskokov and Kroll(2008)]{Goloskokov:2007nt}
S.~V. Goloskokov and P.~Kroll.
\newblock {The Role of the quark and gluon GPDs in hard vector-meson electroproduction}.
\newblock \emph{Eur. Phys. J. C}, 53:\penalty0 367--384, 2008.
\newblock \doi{10.1140/epjc/s10052-007-0466-5}.

\bibitem[Hatta et~al.(2025)Hatta, Klest, Passek-K., and Schoenleber]{Hatta:2025vhs}
Yoshitaka Hatta, Henry~T. Klest, Kornelija Passek-K., and Jakob Schoenleber.
\newblock {Deeply virtual $\phi$-meson production near threshold}.
\newblock 1 2025.
\newblock \doi{10.1093/ptep/ptaf076}.

\bibitem[Klest(2025)]{Klest:2025yik}
Henry~T. Klest.
\newblock {Studying Two-Photon Exchange in Deep Inelastic Scattering with the HERA Data}.
\newblock 7 2025.

\bibitem[Galanti et~al.(2015)Galanti, Giammanco, Grossman, Kats, Stamou, and Zupan]{Galanti:2015pqa}
Mario Galanti, Andrea Giammanco, Yuval Grossman, Yevgeny Kats, Emmanuel Stamou, and Jure Zupan.
\newblock {Heavy baryons as polarimeters at colliders}.
\newblock \emph{JHEP}, 11:\penalty0 067, 2015.
\newblock \doi{10.1007/JHEP11(2015)067}.

\bibitem[Gelis et~al.(2010)Gelis, Iancu, Jalilian-Marian, and Venugopalan]{Gelis:2010nm}
Francois Gelis, Edmond Iancu, Jamal Jalilian-Marian, and Raju Venugopalan.
\newblock {The Color Glass Condensate}.
\newblock \emph{Ann. Rev. Nucl. Part. Sci.}, 60:\penalty0 463--489, 2010.
\newblock \doi{10.1146/annurev.nucl.010909.083629}.

\bibitem[Boussarie et~al.(2016)Boussarie, Grabovsky, Szymanowski, and Wallon]{Boussarie:2016ogo}
R.~Boussarie, A.~V. Grabovsky, L.~Szymanowski, and S.~Wallon.
\newblock {On the one loop $ {\gamma}^{\left(\ast \right)}\to q\overline{q} $ impact factor and the exclusive diffractive cross sections for the production of two or three jets}.
\newblock \emph{JHEP}, 11:\penalty0 149, 2016.
\newblock \doi{10.1007/JHEP11(2016)149}.

\bibitem[Tornqvist(1981)]{Tornqvist:1980af}
Nils~A. Tornqvist.
\newblock {Suggestion for Einstein-podolsky-rosen Experiments Using Reactions Like $e^+ e^- \to \Lambda \bar{\Lambda} \to \pi^- p \pi^+ \bar{p}$}.
\newblock \emph{Found. Phys.}, 11:\penalty0 171--177, 1981.
\newblock \doi{10.1007/BF00715204}.

\bibitem[Lin et~al.(2025)Lin, Liu, Shao, and Wei]{Lin:2025eci}
Shi-Jia Lin, Ming-Jun Liu, Ding~Yu Shao, and Shu-Yi Wei.
\newblock {Spin correlations and Bell nonlocality in $\Lambda\bar{\Lambda}$ pair production from $e^+e^-$ collisions with a thrust cut}.
\newblock 7 2025.

\bibitem[Gu et~al.(2025)Gu, Lin, Shao, Wang, and Yang]{Gu:2025ijz}
Jiayin Gu, Shi-Jia Lin, Ding~Yu Shao, Lian-Tao Wang, and Si-Xiang Yang.
\newblock {Decoherence in high energy collisions as renormalization group flow}.
\newblock 10 2025.

\end{thebibliography}

\bibliographystyle{unsrtnat}
\end{document}